\documentclass[%
 reprint,
 amsmath,amssymb,
 aps,
 prd,
]{revtex4-2}

\usepackage{graphicx}
\usepackage{dcolumn}
\usepackage{bm}
\usepackage{comment}
\usepackage[colorlinks=true,linkcolor=blue,citecolor=blue,urlcolor=blue]{hyperref}

\newcommand{\fft}[2]{\frac{#1}{#2}}
\newcommand{\nn}{\nonumber}
\newcommand{\ri}{i}
\newcommand{\Tr}{\mathrm{Tr}}
\newcommand{\mA}{\mathcal{A}}
\newcommand{\mB}{\mathcal{B}}
\newcommand{\mC}{\mathcal{C}}
\newcommand{\mI}{\mathcal{I}}
\newcommand{\mV}{\mathcal{V}}
\newcommand{\mO}{\mathcal{O}}
\newcommand{\mR}{\mathcal{R}}
\newcommand{\mW}{\mathcal{W}}
\newcommand{\mZ}{\mathcal{Z}}
\newcommand{\mN}{\mathcal{N}}
\newcommand{\mfg}{\mathfrak{g}}

\newcommand{\mm}{\mathfrak{m}}
\newcommand{\mn}{\mathfrak{n}}
\newcommand{\mkt}{\mathfrak{t}}
\newcommand{\tx}{\tilde{x}}
\newcommand{\tu}{\tilde{u}}

\newcommand{\tD}{\widetilde{\Delta}}
\newcommand{\tmn}{\widetilde{\mathfrak{n}}}
\newcommand{\tpsi}{\widetilde{\psi}}

\newcommand{\trho}{\widetilde{\rho}}
\newcommand{\boldDelta}{\bm{\Delta}}
\newcommand{\boldmn}{\boldsymbol{\mn}}
\newcommand{\boldvarphi}{\bm{\varphi}}
\newcommand{\boldtDelta}{\widetilde{\bm{\Delta}}}
\newcommand{\boldtmn}{\widetilde{\bm{\mathfrak{n}}}}
\newcommand{\boldk}{\bm{k}}
\newcommand{\boldm}{\bm{\mathfrak{m}}}
\newcommand{\boldh}{\bm{h}}
\newcommand{\boldu}{\bm{u}}
\newcommand{\boldz}{\bm{z}}
\newcommand{\boldy}{\bm{y}}
\newcommand{\boldxi}{\bm{\xi}}
\newcommand{\boldhu}{\hat{\bm{u}}}
\newcommand{\boldhv}{\hat{\bm{v}}}
\newcommand{\Dsc}{\boldDelta_{\rm sc}}
\newcommand{\tDsc}{\boldtDelta_{\rm sc}}
\newcommand{\Nf}{N_{\rm f}}

\makeatletter
\newif\ifM@firstsec \M@firstsectrue
\newcommand{\M@secskip}{\par
	\ifM@firstsec\global\M@firstsecfalse\else
		\hrule \@height 0.55cm \@width \z@\relax
	\fi}
\renewcommand\section{\M@secskip\@startsection{section}{1}{\z@}%
	{0.25cm \@plus 1ex \@minus .2ex}{0.5cm}{\normalfont\small\bfseries\centering}}
\renewcommand\subsection{\M@secskip\@startsection{subsection}{2}{\z@}%
	{0.25cm \@plus 1ex \@minus .2ex}{0.5cm}{\normalfont\small\bfseries\centering}}
\renewcommand\subsubsection{\M@secskip\@startsection{subsubsection}{3}{\z@}%
	{0.25cm \@plus 1ex \@minus .2ex}{0.5cm}{\normalfont\small\itshape\centering}}
\makeatother

\begin{document}

\preprint{}

\title{Constants in Sequences of M2-brane Partition Functions}

\author{Junho Hong}
\email{junhohong@sogang.ac.kr}
\affiliation{%
 Department of Physics \& Center for Quantum Spacetime, Sogang University,\\
 35 Baekbeom-ro, Mapo-gu, Seoul 04107, Republic of Korea
}%

\date{\today}

\begin{abstract}
We determine in closed form the $N$-independent constant terms in the all-order $1/N$ expansions of the topologically twisted index and of the associated Bethe potential for the ABJM theory, as well as the Bethe potential constant of the ADHM theory. We reconstruct these constants analytically from high-precision Bethe-Ansatz numerics, verify them down to the level of non-perturbative corrections, and find them to be closely related to the constant map function $A$ governing the round three-sphere partition function. The resulting expressions for the ADHM and ABJM constants pass the non-trivial test dictated by 3d mirror symmetry. Via recently established factorization relations, they also determine in closed form the $N$-independent constant contribution to the squashed three-sphere partition function, through the first two leading orders in its large-squashing expansion. These constants supply precisely the piece left undetermined in the recent exact results for 3d supersymmetric partition functions to all orders in the $1/N$ expansion, and thereby mark an important step toward completing them, both in field theory and in the dual quantum gravity description.
\end{abstract}

\maketitle

\section{Introduction}\label{sec:intro}

Supersymmetric localization \cite{Pestun:2007rz,Pestun:2016zxk} has developed into a tool of remarkable precision for holography. It reduces the partition functions of supersymmetric QFTs on compact Euclidean manifolds to finite-dimensional matrix models, enabling exact results that probe string/M-theory beyond the leading supergravity approximation --- most notably the microscopic counting of supersymmetric AdS black hole entropy, see \cite{Benini:2015eyy,Choi:2018hmj,Cabo-Bizet:2018ehj,Benini:2018ywd} for example.

For a large class of 3d holographic SCFTs living on $N$ coincident M2-branes at the tip of a cone over a Sasaki-Einstein 7-manifold, this program has recently reached \emph{all} orders in the $1/N$ expansion across a family of supersymmetric partition functions. The squashed sphere partition function \cite{Kapustin:2009kz,Jafferis:2010un,Hama:2010av,Hama:2011ea,Imamura:2011wg} of this class is conjectured to resum into a compact Airy form \cite{Bobev:2022jte,Bobev:2022eus,Bobev:2023lkx,Bobev:2025ltz} (see also \cite{Hristov:2021qsw,Hristov:2022lcw}), generalizing the original ABJM round sphere results \cite{Fuji:2011km,Marino:2011eh}; the topologically twisted index (TTI) \cite{Benini:2015noa,Benini:2016hjo} and its Seifert generalizations \cite{Closset:2016arn,Closset:2017zgf,Closset:2018ghr,Closset:2019hyt}, which count the microstates of static BPS black holes in AdS$_4$ \cite{Benini:2015eyy}, admit closed all-order expansions \cite{Bobev:2022jte,Bobev:2022eus,Bobev:2023lkx,Hong:2024uns}; the superconformal index (SCI) \cite{Bhattacharya:2008zy,Kim:2009wb,Imamura:2011su} allows for a similar all-order $1/N$ expansion at the first two leading orders of its Cardy-like expansion \cite{Bobev:2022wem,Bobev:2024mqw}; and factorization relations intertwine all such supersymmetric partition functions \cite{Bobev:2026lvl} in the spirit of the holomorphic block decomposition \cite{Pasquetti:2011fj,Beem:2012mb,Hwang:2012jh}. The up-to-date field theory results in this direction are well encapsulated in \cite{Bobev:2025ltz,Bobev:2026lvl}.

On the gravity side, a series of complementary approaches has been steadily closing in on a first-principles understanding of how this all-order structure emerges. The Airy form of the ABJM free energy was reexamined at finite $N$ from higher-derivative 4d gauged supergravity organized into Nekrasov-like gravitational blocks \cite{Hristov:2021qsw,Hristov:2022lcw,Hristov:2024cgj}; equivariant topological strings on Calabi-Yau four-folds successfully geometrized the data governing the Airy formula \cite{Cassia:2025aus,Cassia:2025jkr}; equivariant localization in supergravity \cite{BenettiGenolini:2023kxp,BenettiGenolini:2024xeo,BenettiGenolini:2026qdm} has recently culminated in an Airy representation of the quantum M-theory path integral as well \cite{BenettiGenolini:2026cyc}.

One piece of this picture remains conspicuously less understood: the $N$-independent constant term of the free energies $F=-\log Z$. On the field theory side it is known in closed-form essentially only for special cases of the squashed sphere partition function: the round sphere ABJM constant is the constant map function $A$ of \eqref{Afct} below, obtained through a delicate chain of dualities relating the ABJM matrix model to topological strings or to a free Fermi gas system \cite{Marino:2009jd,Drukker:2010nc,Fuji:2011km,Marino:2011eh,Hanada:2012si,Hatsuda:2014vsa,Grassi:2014zfa}, and the closed-forms at special squashing and mass parameters rest on similar dualities \cite{Nosaka:2015iiw,Hatsuda:2016uqa,Kubo:2024qhq,Kubo:2025dot}. Away from these special cases, the constants have been available only numerically \cite{Bobev:2022jte,Bobev:2022eus,Bobev:2022wem,Bobev:2023lkx,Bobev:2025ltz}. The gravity-side derivations above do not fix it either: they reproduce the Airy data but in general leave the constant undetermined, with no eleven-dimensional derivation of the constant map contribution currently available \cite{Bobev:2026gir}. 
Yet the known special cases are strikingly structured: for the ABJM round sphere, the $N$-independent constant is precisely the all-genus sum of the constant map contributions of the dual topological string on local $\mathbb P^1\times\mathbb P^1$ \cite{Aganagic:2002wv,Marino:2009jd,Drukker:2010nc,Drukker:2011zy}, resummed into the compact function $A$ as reviewed below \cite{Marino:2011eh,Hanada:2012si,Hatsuda:2012dt,Hatsuda:2014vsa}. Such closed-form access to constant contributions in more general M2-brane worldvolume theories on various supersymmetric backgrounds would thus bring the $N$-independent term to the same universal footing that the $N$-dependent data now enjoy.

In this work we fill in this gap, focusing on the TTI and its associated Bethe potential in the ABJM and ADHM theories. Section~\ref{sec:setup} reviews the localization formulae for the TTI, the SCI and the squashed sphere partition function, and introduces the constant map function $A$. Section~\ref{sec:ABJM} determines the ABJM Bethe potential and TTI constants at the superconformal point in closed-form, and obtains partial results at generic flavor chemical potentials; all of them turn out to be finite linear combinations of the constant map function with direct and inverted arguments, deduced from high-precision numerical data with the valuable help of the AI assistant Claude Fable 5 in the pattern-recognition steps. Section~\ref{sec:ADHM} establishes the parallel ADHM results, including an exact relation between the ADHM and ABJM Bethe potential constants dictated by 3d mirror symmetry. Section~\ref{sec:Airy} leverages these closed-forms, through the factorization relations of \cite{Bobev:2026lvl}, to determine the first two leading orders in the large-squashing expansion of the Airy constant of the squashed sphere partition function, and assembles the enlarged web of exact constraints toward its complete determination. Section~\ref{sec:discussion} concludes with open questions, and two appendices provide computational details behind the main text.

\medskip
\noindent\emph{Note added.} While this work was nearing completion, we became aware of the independent work \cite{Hosseini:2026}, which addresses closely related questions and has partial overlap with the results presented here.

\section{Localization formulae for 3d supersymmetric partition functions}\label{sec:setup}

In this section we briefly review the localization formulae of three different supersymmetric partition functions for 3d $\mathcal N=2$ Chern-Simons(CS)-matter quiver gauge theories that may flow to various holographic SCFTs of interest in the IR. Throughout we employ the collective notation summarized below. 
\begin{itemize}
	\item \emph{Symmetry.} We denote by $G$ the gauge group, of rank $r_G$ and Lie algebra $\mfg$, and by $F$ the global symmetry group of rank $r_F$. The symbol $\alpha$ denotes the roots of $\mfg$, forming the set $\mR[\mfg]$. The Weyl group of $G$ is denoted $\mW$.
	\item \emph{Matter.} The matter sector is built from $\mathcal N=2$ chiral multiplets, each denoted $\Psi$ and assigned an $R$-charge $r_\Psi$. Every $\Psi$ is in a specific representation of $G\times F$, and we write $\rho$ and $\trho$ for the weights of its gauge and flavor representations. 
	\item \emph{Gauge zero modes and fluxes.} The constant modes of localization residing in $\mN=2$ gauge vector multiplets are integrated over, and correspond to the gauge holonomies $\boldh$ on $S^1\times S^2$ and the scalar constant modes $\boldhu$ on $S^3_b$. On $S^1\times S^2$ one sums in addition over the gauge magnetic fluxes $\boldm$ through the $S^2$, whose quantization places them in the co-root lattice; throughout this paper we take them integer-valued, identifying $Q^\vee(\mfg)\simeq\mathbb Z^{r_G}$. A boldface symbol always stands for the full collection of its components, so that for instance $d\boldhu\equiv\prod_{\ell=1}^{r_G}d\hat u_\ell$.
	\item \emph{Background data.}  The constant modes in background vector multiplets are kept fixed and characterize the supersymmetric partition function. They act as real masses $\boldhv$ on $S^3$ \cite{Freedman:2013oja}, while we combine them with background gauge holonomies on $S^1\times S^2$ and write flavor chemical potentials $\boldDelta$ \cite{Benini:2015noa}, whose exponentials correspond to the flavor fugacities $\boldxi$. The associated flavor magnetic fluxes are written as $\boldmn$. 
	\item \emph{Chern-Simons levels.} The single symbol $\boldk$ gathers the three types of CS levels that a theory in this class may carry: $k^{\ell n}$ between two dynamical gauge fields, $k^{xy}$ between two background fields, and $k^{\ell x}$ mixing the two. 
\end{itemize}
Here and below we use the same symbols $\mZ_{\rm CS}$, $\mZ_{\rm VM}$ and $\mZ_{\rm CM}$ for the classical CS, vector multiplet and chiral multiplet contributions to each of the three supersymmetric partition functions.

\subsection{Topologically twisted index}\label{sec:setup:TTI}
The TTI is the supersymmetric partition function on $S^1\times\Sigma_{g}$ with a partial topological twist along the Riemann surface $\Sigma_{g}$ \cite{Benini:2015noa,Benini:2016hjo,Closset:2016arn}, where we focus on the special case $\Sigma_{g}=S^2~(g=0)$ in this paper. Supersymmetric localization reduces it to a matrix model as \cite{Benini:2015noa,Benini:2015eyy,Hosseini:2016tor}
\begin{align}
	Z_{S^1\times S^2}(\boldk,\boldDelta,\boldmn)&=\fft{1}{|\mW|}\sum_{\boldm\in Q^\vee(\mfg)}\oint_{\rm JK}\fft{d\boldh}{(2\pi)^{r_G}}\nn\\
	&\quad\times\mZ_{\rm CS}\,\mZ_{\rm VM}\,\mZ_{\rm CM}\,,\label{TTI:general}
\end{align}
where the subscript JK indicates that the contour picks up the Jeffrey-Kirwan residues \cite{Benini:2015noa} and the three building blocks read 
\begin{align}
	\mZ_{\rm CS}&=\prod_{\ell=1}^{r_G}z_\ell^{\sum_nk^{\ell n}\mm_n+\sum_xk^{\ell x}\mn_x}\prod_{x=1}^{r_F}\xi_x^{\sum_yk^{xy}\mn_y+\sum_\ell k^{x\ell}\mm_\ell}\,,\nn\\
	\mZ_{\rm VM}&=\prod_{\alpha\in\mR[\mfg]}\big(1-\boldz^\alpha\big)\,,\nn\\
	\mZ_{\rm CM}&=\prod_\Psi\prod_{\rho,\trho}\bigg(\fft{\sqrt{\boldz^\rho\boldy^{\trho}}}{1-\boldz^\rho\boldy^{\trho}}\bigg)^{\rho(\boldm)-\trho(\boldmn)+1}\,.\label{TTI:blocks}
\end{align}
Here the flavor chemical potentials and magnetic fluxes, $\boldDelta$ and $\boldmn$, are constrained by the marginality of the superpotential under the global symmetries and by supersymmetry. In $\mZ_{\rm CS}$ the mixed CS levels generate in particular the topological symmetry factors of a quiver gauge theory.

The JK residues in \eqref{TTI:general} can be summed into the Bethe-Ansatz (BA) formula as \cite{Benini:2015noa,Benini:2015eyy,Closset:2017zgf}
\begin{align}
	Z_{S^1\times S^2}=\sum_{\boldu\,\in\,\mathrm{BAE}}\mZ_{\rm int}(\boldu)\det\mathbb B(\boldu)^{-1}\,,\label{BA:general}
\end{align}
where the sum runs over the solutions of the Bethe-Ansatz equations (BAE), $\mZ_{\rm int}$ is obtained from the integrand of \eqref{TTI:general} by summing over the gauge magnetic fluxes, and $\mathbb B$ is the Jacobian of the BAE with respect to the gauge zero modes $\boldu$. The BAE themselves take the form
\begin{align}
	e^{\ri B_\ell(\boldu)}=1\qquad\text{where}\qquad B_\ell\equiv\fft{\partial\mV}{\partial u_\ell}\,,\label{BAE:general}
\end{align}
i.e.\ they are the extremization conditions (modulo $2\pi\mathbb Z$) of the Bethe potential \cite{Benini:2015eyy}, which also plays the role of an effective twisted superpotential governing the low-energy
dynamics on the Coulomb branch of the $A$-twist theory \cite{Closset:2016arn,Closset:2017zgf}. We omit its general expression here in consideration of subtle convention differences across the literature \cite{Benini:2015eyy,Closset:2016arn,Closset:2017zgf,Hosseini:2016tor,PandoZayas:2020iqr,Bobev:2023lkx} as discussed in \cite{Bobev:2024mqw,Hong:2024uns}. Instead we present the explicit Bethe potentials $\mV$ for the two examples in Sections \ref{sec:ABJM} and \ref{sec:ADHM}, as well as the corresponding residue expressions $\mZ_{\rm int}$ determining the TTI $Z_{S^1\times S^2}$ through the BA formula \eqref{BA:general}. The main interest of this paper lies in the $N$-independent constant contributions to the $1/N$ expansions of the Bethe potential and the TTI for those two examples.

\subsection{Superconformal index}\label{sec:setup:SCI}
The SCI is the supersymmetric partition function on $S^1\times_\omega S^2$ counting local BPS operators, defined as the trace over the Hilbert space of the theory in radial quantization \cite{Bhattacharya:2008zy,Bhattacharya:2008bja,Kim:2009wb}. Supersymmetric localization yields the matrix model representation \cite{Imamura:2011su,Krattenthaler:2011da,Kapustin:2011jm}
\begin{align}
	\mI(q,\boldxi,\boldmn)=\fft{1}{|\mW|}\sum_{\boldm\in Q^\vee(\mfg)}\int_0^{2\pi}\fft{d\boldh}{(2\pi)^{r_G}}\;\mZ_{\rm CS}\mZ_{\rm VM}\mZ_{\rm CM}\,,\label{SCI:general}
\end{align}
where the building blocks read
\begin{align}
	\mZ_{\rm VM}&=\prod_{\alpha\in\mR[\mfg]}(qe^{\pi i})^{-\fft12|\alpha(\boldm)|}\Big(1-\boldz^\alpha q^{|\alpha(\boldm)|}\Big)\,,\label{SCI:blocks}\\
	\mZ_{\rm CM}&=\prod_\Psi\prod_{\rho,\trho}\Big(q^{1-r_\Psi}\boldz^{-\rho}\boldxi^{-\trho}e^{\pi i(1-r_\Psi)}\Big)^{\fft12|\rho(\boldm)+\trho(\boldmn)|}\nn\\
	&\kern4em\times\fft{\big(e^{-\pi ir_\Psi}\boldz^{-\rho}\boldxi^{-\trho}q^{2-r_\Psi+|\rho(\boldm)+\trho(\boldmn)|};q^2\big)_\infty}{\big(e^{\pi ir_\Psi}\boldz^{\rho}\boldxi^{\trho}q^{r_\Psi+|\rho(\boldm)+\trho(\boldmn)|};q^2\big)_\infty}\,,\nn
\end{align}
with the classical CS contribution the same as in \eqref{TTI:blocks} and $(\cdot\,;\cdot)_\infty$ the $\infty$-Pochhammer symbol. Subtle phase corrections arising in the localization formula are associated with the grading factor in the trace formula, see \cite{Aharony:2013dha,Bobev:2022wem,Bobev:2024mqw,Bobev:2026lvl} for related discussions.

In the Cardy-like limit $\omega\to\ri0^+$ with $q=e^{\pi i \omega}$, the SCI can be evaluated by a saddle point approximation \cite{Choi:2019zpz,Choi:2019dfu,Nian:2019pxj}. On the vanishing flavor magnetic flux section with $\boldmn=0$, it turns out that the corresponding saddle point equations are directly identified with the TTI BAEs \eqref{BAE:general} for a large class of $\mN=2$ CS-matter theories \cite{Bobev:2022wem,Bobev:2024mqw}. The first two orders of the Cardy-like expansion of the SCI are then governed by the Bethe potential and by the TTI respectively, through the key relation \cite{Bobev:2022wem,Bobev:2024mqw}
\begin{align}
	\log\mI
	=-\fft{1}{\pi\omega}\,\Im\mV
	+\Re\big[\log Z_{S^1\times S^2}\big]+\ri\vartheta+\mO(\omega)\,,\label{SCI:Cardy}
\end{align}
where the Bethe potential and the TTI on the right-hand side are evaluated on the contributing Bethe vacua and $\vartheta$ is a real phase independent of the details of the theory. We refer the reader to \cite{Bobev:2022wem,Bobev:2024mqw} for the precise map between the flavor parameters of the two indices. 

We do not investigate the SCI directly in this paper; the relation \eqref{SCI:Cardy}, however, immediately transfers the new results of Sections~\ref{sec:ABJM} and \ref{sec:ADHM} for the $N$-independent constants of the Bethe potential and of the TTI to the corresponding constants in the Cardy-like expansion of the SCI.

\subsection{Squashed sphere partition function}\label{sec:setup:S3b}
The partition function on the $\mathrm{U}(1)\times\mathrm{U}(1)$ invariant squashed sphere $S^3_b$ with squashing parameter $b$ localizes to a matrix model built out of double sine functions $s_b$ \cite{Hama:2011ea,Imamura:2011wg}, in close analogy with the round sphere case \cite{Kapustin:2009kz,Kapustin:2010xq,Jafferis:2010un,Hama:2010av}. The resulting matrix model reads 
\begin{align}
	Z(b,\boldk,\boldhv)=\fft{1}{|\mW|}\int d\boldhu\;\mZ_{\rm CS}\,\mZ_{\rm VM}\,\mZ_{\rm CM}\,,\label{S3b:general}
\end{align}
with $Q\equiv b+b^{-1}$ and
\begin{align}
	\mZ_{\rm CS}&=e^{-\pi\ri\left[\sum_{\ell,n}k^{\ell n}\hat u_\ell\hat u_n+\sum_{x,y}k^{xy}\hat v_x\hat v_y+2\sum_{\ell,x}k^{\ell x}\hat u_\ell\hat v_x\right]}\,,\nn\\
	\mZ_{\rm VM}&=\prod_{\alpha\in\mR[\mfg]}s_b\bigg(\fft{\ri Q}{2}-\alpha(\boldhu)\bigg)^{-1}\,,\nn\\
	\mZ_{\rm CM}&=\prod_\Psi\prod_{\rho,\trho}s_b\bigg(\fft{\ri Q}{2}(1-r_\Psi)-\rho(\boldhu)-\trho(\boldhv)\bigg)\,.\label{S3b:blocks}
\end{align}

Recently, for a large class of $\mN=2$ CS-matter quiver gauge theories characterized by a product unitary gauge group $\otimes_{r=1}^p\mathrm{U}(N)_{k_r}$ with vanishing total CS level and non-chiral matter content, the $S^3_b$ partition function \eqref{S3b:general} is conjectured to take the Airy form \cite{Bobev:2025ltz}
\begin{align}
	Z=\mC^{-1/3}e^{\mA}\,\mathrm{Ai}\big[\mC^{-1/3}(N-\mB)\big]\big(1+\mO(e^{-\#\sqrt N})\big)\label{Airy}
\end{align}
in the $1/N$ expansion up to non-perturbative corrections. The conjecture is established for the round sphere ABJM theory in \cite{Fuji:2011km,Marino:2011eh} and in a number of special cases with mass and squashing deformations \cite{Nosaka:2015iiw,Hatsuda:2016uqa,Hatsuda:2021oxa,Chester:2023qwo,Kubo:2024qhq} as summarized in \cite{Bobev:2025ltz}. The coefficients $\mB$ and $\mC$ are known in closed-form for various low energy worldvolume theories of $N$ coincident M2-branes probing a cone over a seven-dimensional Sasaki-Einstein manifold, which are therefore holographically dual to M-theory on the corresponding AdS$_4$ backgrounds. They take the universal form 
\begin{align}
	\mB=\beta-\fft{4}{3Q^2}\gamma\qquad\&\qquad
	\mC=\bigg(\fft{8}{3\pi Q^2\alpha}\bigg)^2\label{Airy:BC}
\end{align}
so that the squashing enters $\mB$ and $\mC$ only through $Q^2$. Here $(\alpha,\beta,\gamma)$ are three theory-dependent functions of the CS levels and the flavor parameters, whose explicit values for the various holographic SCFTs are recorded in \cite{Bobev:2025ltz}.  

In contrast, a closed-form expression for the $N$-independent constant $\mA$ is not known in general; the special cases where it is known are collected in \cite{Bobev:2025ltz} and the ones we need are recalled in Section~\ref{sec:Airy:anchors}. A central role in all such special cases --- and, as we will see, in the $N$-independent constants of the Bethe potential and of the TTI as well --- is played by the constant map function $A$, defined through the representations \cite{Marino:2011eh,Hanada:2012si,Hatsuda:2012dt,Hatsuda:2014vsa}
\begin{align}
	A(k)&\equiv\fft{2\zeta(3)}{\pi^2k}\bigg(1-\fft{k^3}{16}\bigg)+\fft{k^2}{\pi^2}\int_0^\infty dx\,\fft{x\,\log\big(1-e^{-2x}\big)}{e^{kx}-1}\nn\\
	&=-\fft{\zeta(3)}{8\pi^2}k^2+2\zeta'(-1)+\fft16\log\fft{4\pi}{k}\nn\\
	&\quad+\sum_{n=1}^{\infty}\bigg(\fft{2\pi}{k}\bigg)^{2n}\fft{(-1)^{n+1}4^n|B_{2n}B_{2n+2}|}{(n+1)\,2n\,(2n)!}\nn\\
	&=\fft{2\zeta(3)}{\pi^2k}-\fft{k}{12}-\fft{\pi^2k^3}{4320}+\fft{\pi^4k^5}{907200}-\ldots\,,\label{Afct}
\end{align}
where the second representation is the asymptotic expansion at large $k$ in terms of Bernoulli numbers and the third is the convergent expansion at small $k$. The exact values at all integer arguments are known \cite{Hatsuda:2014vsa} and we list some examples below for later use:
\begin{align}
	A(1)&=-\fft{\zeta(3)}{8\pi^2}+\fft{\log2}{4}\,,&~\,
	A(2)&=-\fft{\zeta(3)}{2\pi^2}\,,\label{Afct:special}\\
	A(4)&=-\fft{\zeta(3)}{4\pi^2}-\fft{\log2}{2}\,,&
	A(8)&=-\fft{\zeta(3)}{8\pi^2}-\fft74\log2\,.\nn
\end{align}
The integral representation can be employed to prove that $A(k)$ is an odd function of its argument as $A(-k)=-A(k)$ \cite{Bobev:2025ltz}, a property that will be used repeatedly below. In the topological string interpretation of the round sphere partition function, $A(k)$ originates from the all-genus resummation of constant-map contributions \cite{Bershadsky:1993cx,Drukker:2010nc,Drukker:2011zy,Hanada:2012si,Hatsuda:2012dt,Hatsuda:2014vsa}; its ubiquitous appearance in the $N$-independent constants of M2-brane partition functions beyond the known duality chain is a main theme of this paper.

The squashed sphere partition function is moreover tied to the two indices reviewed above by factorization. As observed in the Cardy-like limit \cite{Choi:2019dfu} and further established to all orders \cite{Bobev:2026lvl}, both the SCI \eqref{SCI:general} and the TTI \eqref{TTI:general} factorize into a product of two copies of \eqref{S3b:general}, evaluated at squashings and flavor parameters determined by the corresponding index variables. At the level of the $N$-independent constants, this turns the Airy constant $\mA$ into a quantity that can be read off from the $N$-independent constants of the Bethe potential and of the TTI, and conversely; we spell out the relations we need, and their consequences, in Section~\ref{sec:Airy}.

\medskip

Concluding this review section, we emphasize again that once our primary objects of study in the next sections --- the $N$-independent constants of the Bethe potential and the TTI --- are known in closed-form, the Cardy-like relation \eqref{SCI:Cardy} immediately supplies the corresponding constants of the SCI, and the factorization relations just mentioned supply those of the squashed sphere partition function, i.e.\ the Airy constant $\mA$. A single field theory computation therefore propagates through the whole web of supersymmetric partition functions of these theories.

\section{Constants of the ABJM theory}\label{sec:ABJM}

In this section we determine the $N$-independent constants in the all-order $1/N$ expansions of the ABJM Bethe potential and TTI. We first briefly introduce the ABJM theory, specialize the localization formulae of Section~\ref{sec:setup} to it, and review the known all-order results that define these constants. We then present our new results.

\subsection{Review}\label{sec:ABJM:review}
The ABJM theory is the 3d $\mathcal N=6$ $\mathrm{U}(N)_k\times\mathrm{U}(N)_{-k}$ CS-matter theory describing the low-energy dynamics of $N$ M2-branes probing a $\mathbb C^4/\mathbb Z_k$ singularity, holographically dual to M-theory on AdS$_4\times S^7/\mathbb Z_k$ \cite{Aharony:2008ug}. In $\mathcal N=2$ language, its quiver consists of two $\mathrm{U}(N)$ nodes with opposite CS levels $\pm k$ and two pairs of bi-fundamental and anti-bi-fundamental chiral multiplets $A_{1,2}$ and $B_{1,2}$ with superconformal R-charge $\fft12$, interacting through the quartic superpotential
\begin{align}
	W\propto\Tr\big[A_1B_1A_2B_2-A_1B_2A_2B_1\big]\,.\label{ABJM:W}
\end{align}
The global symmetry manifest in this $\mathcal N=2$ formulation is $\mathrm{SU}(2)_A\times\mathrm{SU}(2)_B\times\mathrm{U}(1)_T\times\mathrm{U}(1)_R$, where the two $\mathrm{SU}(2)$ factors rotate $A_{1,2}$ and $B_{1,2}$ respectively and $\mathrm{U}(1)_T$ is the topological symmetry.

In the localization formula of the ABJM TTI, the Cartan of the global symmetry is encoded in four flavor chemical potentials $\Delta_a$ entering through the fugacities $y_a=e^{\ri\pi\Delta_a}$ and four flavor magnetic fluxes $\mn_a$ $(a=1,\ldots,4)$, subject to the constraints following from the marginality of the superpotential \eqref{ABJM:W} and from supersymmetry 
\begin{align}
	\sum_{a=1}^4\Delta_a=2\,,\qquad \sum_{a=1}^4\mn_a=2\,.\label{ABJM:constraints}
\end{align}
The superconformal point corresponds to the configuration
\begin{align}
	\boldDelta_{\rm sc}:~\Delta_a=\fft12\,,\qquad \boldmn_{\rm sc}:~\mn_a=\fft12\,.\label{sc:config}
\end{align}
For the ABJM theory, the BA formula \eqref{BA:general} takes the explicit form 
\begin{widetext}
\begin{align}
	Z_{S^1\times S^2}(N,k,\boldDelta,\boldmn)
	=\prod_{a=1}^4y_a^{-\fft{N^2}{2}\mn_a}\sum_{\{x_i,\tx_j\}\in\mathrm{BAE}}\fft{1}{\det\mathbb B}\,\fft{\prod_{i=1}^Nx_i^N\tx_i^N\prod_{i\neq j}^N\big(1-\fft{x_i}{x_j}\big)\big(1-\fft{\tx_i}{\tx_j}\big)}{\prod_{i,j=1}^N\prod_{a=1,2}(\tx_j-x_iy_a)^{1-\mn_a}\prod_{a=3,4}(x_i-\tx_jy_a)^{1-\mn_a}}\,,\label{TTI:BA}
\end{align}
where $x_i=e^{\ri u_i}$ and $\tx_j=e^{\ri\tu_j}$ parametrize the holonomies of the two gauge groups and the BAE Jacobian $\mathbb B$ can be found in \cite{Bobev:2022eus}. The ABJM Bethe potential reads 
\begin{align}
	\mV(N,k,u,\tu,\boldDelta)&=\sum_{i=1}^N\bigg[\fft{k}{2}\big(\tu_i^2-u_i^2\big)-\pi\Big(2\tilde n_i-\fft{1-(-1)^N}{2}\Big)\tu_i+\pi\Big(2n_i-\fft{1-(-1)^N}{2}\Big)u_i\bigg]\nn\\
	&\quad+\sum_{i,j=1}^N\bigg[\sum_{a=3,4}\mathrm{Li}_2\big(e^{\ri(\tu_j-u_i+\pi\Delta_a)}\big)-\sum_{a=1,2}\mathrm{Li}_2\big(e^{\ri(\tu_j-u_i-\pi\Delta_a)}\big)\bigg]\,,\label{TTI:V}
\end{align}
\end{widetext}
where the integers are chosen as $(n_i,\tilde n_i)=(1-i,i-N)$ to obtain the standard large $N$ BAE solution of \cite{Benini:2015eyy} through \eqref{BAE:general}. Let us note again that mild convention differences for these formulae exist across the literature --- possibly arising from phase corrections, branch cut ambiguities, or quantization schemes --- but none of them affects the $N$-independent constants studied in this paper. 

\medskip

The imaginary part of the Bethe potential \eqref{TTI:V} evaluated on the exact BAE solution, which is obtained via numerical analysis based on the analytic large $N$ solution of \cite{Benini:2015eyy}, turns out to take the all-order form \cite{Bobev:2022wem} 
\begin{align}
	\fft{\Im\mV(N,k,\boldDelta)}{2\pi}&=\fft{\pi\sqrt{2k\Delta_1\Delta_2\Delta_3\Delta_4}}{3}\,\hat N_\Delta^{3/2}\nn\\
	&\quad+\hat g_0(k,\boldDelta)+\hat g_{\rm np}(N,k,\boldDelta)\,,\label{ImV}
\end{align}
where the shifted rank $\hat N_\Delta$ is defined as \cite{Bobev:2022eus} 
\begin{align}
	\hat N_\Delta\equiv N-\fft{k}{24}+\fft{1}{12k}\sum_{a=1}^4\fft{1}{\Delta_a}\,,\label{Nhat}
\end{align}
and $\hat g_{\rm np}$ encapsulates non-perturbative corrections. The all-order $1/N$ expansion of the TTI \eqref{TTI:BA} is obtained in a similar manner as \cite{Bobev:2022jte,Bobev:2022eus} 
\begin{align}
	&\log Z_{S^1\times S^2}(N,k,\boldDelta,\boldmn)\nn\\
	&=-\fft{\pi\sqrt{2k\Delta_1\Delta_2\Delta_3\Delta_4}}{3}\sum_{a=1}^4\fft{\mn_a}{\Delta_a}\Big(\hat N_\Delta^{3/2}-\fft{\mathfrak c_a}{k}\hat N_\Delta^{1/2}\Big)\nn\\
	&\quad-\fft12\log\hat N_\Delta+\log k+\hat f_0(k,\boldDelta,\boldmn)\nn\\
	&\quad+\hat f_{\rm np}(N,k,\boldDelta,\boldmn)\,,\label{TTI:allorder}
\end{align}
with the rational coefficients 
\begin{align}
	\mathfrak c_a=\fft{\prod_{b\neq a}(\Delta_a+\Delta_b)}{8\Delta_1\Delta_2\Delta_3\Delta_4}\sum_{b\neq a}\Delta_b\,.\label{ca}
\end{align}
In \eqref{TTI:allorder} we have split off the universal $\log k$ contribution arising from the $k$-fold degeneracy \cite{Benini:2015eyy,Bobev:2022eus}, so that $\hat f_0$ is the $N$-independent constant of a \emph{single} Bethe vacuum. This convention is natural in view of the factorization relation derived in \cite{Bobev:2026lvl}, which isolates a single Bethe vacuum; hence the constant denoted $\hat f_0$ in \cite{Bobev:2022eus} equals $\hat f_0+\log k$ in our conventions. The non-perturbative corrections are of order \cite{Bobev:2022eus} 
\begin{align}
	\hat f_{\rm np}(N,k,\boldDelta,\boldmn)=e^{-2\pi\sqrt{2N/k}+\mO(\log N)}\,,\label{fnp}
\end{align}
whose leading exponent is reminiscent of the semiclassical action of M2-brane instantons wrapping $S^3/\mathbb Z_k$, or equivalently of type IIA worldsheet instantons wrapping $\mathbb{CP}^1\subset\mathbb{CP}^3$ \cite{Cagnazzo:2009zh,Drukker:2010nc,Hatsuda:2012dt}, computed directly from the quantum M2-brane/worldsheet path integral in \cite{Gautason:2023igo,Beccaria:2023ujc}; the original calculation in the context of holographic duality involving the $S^3$ partition function was recently generalized to cover the case of interest \eqref{fnp} and beyond, see \cite{Gautason:2025per,vanMuiden:2026nsp}.

\medskip

Now let us focus on our primary targets, the $N$-independent constants $\hat g_0$ and $\hat f_0$ appearing in \eqref{ImV} and \eqref{TTI:allorder}. Prior to this work, the Bethe potential constant was known only to the leading order as \cite{Bobev:2022wem,Bobev:2025ltz}
\begin{align}
	\hat g_0(k,\boldDelta)&=\hat g_{0,2}(\boldDelta)\,\fft{\zeta(3)}{8\pi^2}k^2+\mO(\log k)\,,\label{g0:k2}\\
	\hat g_{0,2}(\boldDelta)&=\fft{4-\sum_{a=1}^4\Delta_a^2}{4}\nn\\
	&\quad-\bigg(\fft{1}{\Delta_{13}\Delta_{24}}+\fft{1}{\Delta_{14}\Delta_{23}}\bigg)\sum_{a=1}^4\fft{\prod_{b=1}^4\Delta_b}{\Delta_a}\,,\nn
\end{align}
where $\Delta_{ab}\equiv\Delta_a+\Delta_b$. For the TTI constant at the superconformal point, the large-$k$ expansion was determined numerically as \cite{Bobev:2022eus} 
\begin{align}
	\hat f_0(k,\Dsc,\boldmn_{\rm sc})&=-\fft{3\zeta(3)}{8\pi^2}k^2+\fft16\log k+\mathfrak f_0\nn\\
	&\quad+\sum_{n=1}^{\infty}\bigg(\fft{2\pi}{k}\bigg)^{2n}\,\mathfrak f_n\label{f0:series}
\end{align}
with the $\log k$ coefficient shifted by the degeneracy split in \eqref{TTI:allorder}, where the constant $\mathfrak f_0=-2.09684829977578309$ and the fractions
\begin{align}
	\{\mathfrak f_n\}_{n\leq5}=\Big\{&-\fft{2}{45},\,\fft{19}{5670},\,-\fft{41}{42525},\nn\\
	&\;\fft{31}{56700},\,-\fft{964636}{1915538625}\Big\}\label{f2n:known}
\end{align}
were obtained by high-precision fits, with no closed-form available. For generic flavor chemical potentials and fluxes, only the first two orders of the large $k$ expansion of $\hat f_0$ are known as \cite{Bobev:2022eus,Bobev:2026lvl} 
\begin{align}
	\hat f_0(k,\boldDelta,\boldmn)&=-\fft{\zeta(3)}{8\pi^2}k^2\sum_{a=1}^4\hat f_{0,2,a}(\boldDelta)\,\mn_a\nn\\
	&\quad+\fft16\log k+\mO(k^{0})\,,\label{f0:general}
\end{align}
where the four coefficient functions are generated from a single one by permutations of the flavor labels 
\begin{align}
	\hat f_{0,2,1}(\boldDelta)&=\Delta_1+\fft{\Delta_1\Delta_3}{\Delta_{14}}+\fft{\Delta_1\Delta_4}{\Delta_{13}}+\fft{\Delta_1\Delta_4\Delta_{23}}{\Delta_{14}^2}\nn\\
	&\quad+\fft{\Delta_1\Delta_3\Delta_{24}}{\Delta_{13}^2}-\fft{2\Delta_3\Delta_4}{\Delta_{13}\Delta_{14}}-\fft{\Delta_2^2(\Delta_1-\Delta_2)}{\Delta_{23}\Delta_{24}}\nn\\
	&\quad+\fft{\Delta_2\Delta_3\Delta_{14}}{\Delta_{13}\Delta_{23}}+\fft{\Delta_2\Delta_4\Delta_{13}}{\Delta_{14}\Delta_{24}}\,,\label{f0:general:lead}\\
	\hat f_{0,2,2}(\boldDelta)&=\hat f_{0,2,1}(\boldDelta)\big|_{\Delta_1\leftrightarrow\Delta_2}\,,\nn\\
	\hat f_{0,2,3}(\boldDelta)&=\hat f_{0,2,1}(\boldDelta)\big|_{\Delta_1\leftrightarrow\Delta_3,\,\Delta_2\leftrightarrow\Delta_4}\,,\nn\\
	\hat f_{0,2,4}(\boldDelta)&=\hat f_{0,2,1}(\boldDelta)\big|_{\Delta_1\leftrightarrow\Delta_4,\,\Delta_2\leftrightarrow\Delta_3}\,.\nn
\end{align}
The $\log k$ coefficient in \eqref{f0:general} is universal, i.e.\ independent of $\boldDelta$ and $\boldmn$, and is closely related to the one-loop structure of the dual supergravity \cite{Bobev:2023dwx}. 

\subsection{New results}\label{sec:ABJM:new}

Our new results are based on high-precision numerical solutions of the BAE \eqref{BAE:general} for the ABJM theory, obtained by \texttt{FindRoot} in \textsc{Mathematica} at \texttt{WorkingPrecision} 200 following the algorithm of \cite{Bobev:2022eus} (see also \cite{Liu:2017vll,PandoZayas:2019hdb,PandoZayas:2020iqr}) with the leading large-$N$ solution of \cite{Benini:2015eyy} as the initial condition. In particular, to analyze the large $k$ expansion of the constant terms efficiently, we generate numerical BAE solutions in the type IIA regime of fixed 't~Hooft coupling $\lambda\equiv N/k=30$ with $N=101\sim501$ in steps of $10$ for the Bethe potential $\mV$ and $N=101\sim551$ for the TTI, while a smaller upper bound (\emph{e.g.} $N\leq401$) is taken for different generic flavor configurations drawn from the sets used in Appendix~D of \cite{Bobev:2022eus}, Appendix~C.2 of \cite{Bobev:2022wem}, or Appendix E.2 of \cite{Bobev:2025ltz}. The Bethe potential and the TTI are then evaluated on the BAE solutions, while for the TTI we set the fluxes to $\mn_a=\Delta_a$ for simplicity. The $N$-independent constants are then extracted by subtracting the known $N$-dependent terms of \eqref{ImV} and \eqref{TTI:allorder} from the exact numerical data of $\mV$ and TTI, and fitting the residual via \texttt{LinearModelFit} over the basis $\{k^2,\log k,1,k^{-2n}\}$ with a suitable upper bound on the integer $n$; individual tail coefficients are confirmed order by order by subtracting all lower-order terms exactly before refitting.

\subsubsection{The Bethe potential constant}\label{sec:ABJM:g0}

\emph{Superconformal point.} Our main result for the Bethe potential constant at the superconformal point is the closed-form expression
\begin{align}
	\hat g_0(k,\Dsc)&=-\fft{\zeta(3)}{32\pi^2}k^2-\fft{\log2}{6}\label{g0:closed}\\
	&\quad+\fft{k}{\pi^2}\int_0^\infty\!dx\log(1-e^{-kx})\log\cosh x\nn\\
	&=-\fft{\zeta(3)}{32\pi^2}k^2-\fft{\log2}{6}\nn\\
	&\quad+\sum_{n=1}^{\infty}\bigg(\fft{2\pi}{k}\bigg)^{2n}\fft{(-1)^n4^n(4^n-1)|B_{2n}B_{2n+2}|}{n\,(2n+2)!}\,,\nn
\end{align}
where the second equality is the asymptotic expansion of the integral representation at large $k$, derived in Appendix~\ref{app:series}. 

The expression \eqref{g0:closed} was obtained in the following steps. First, the large-$k$ fits described above determine the tail coefficients of $\hat g_0(k,\Dsc)$ as exact rationals,
\begin{align}
	-\fft{1}{360}\,,\quad\fft{1}{7560}\,,\quad-\fft{1}{37800}\,,\quad\fft{17}{1496880}\,,~\ldots\label{g0:tail:num}
\end{align}
at orders $(2\pi/k)^2,\ldots,(2\pi/k)^8$. Second, motivated by the higher-order structure of the large-$k$ expansion \eqref{Afct} of the constant map function $A$, whose coefficients are built from products of two Bernoulli numbers, we deduced from these rationals the general-order formula in the second line of \eqref{g0:closed} --- a pattern-recognition step in which the AI assistant Claude Fable 5 provided valuable help --- and the predicted exact values $-\fft{21421}{2554051500}$ and $\fft{691}{72972900}$ at orders $(2\pi/k)^{10}$ and $(2\pi/k)^{12}$ were successfully matched with numerical estimation via \texttt{LinearModelFit}. Third, running the derivation of the large-$k$ expansion of $A(k)$ from its integral representation \eqref{Afct} backward, we resummed the general-order series into the integral representation in the first line of \eqref{g0:closed}. Finally, the fitted constant term was identified as $-\fft{\log2}{6}$ within its precision, motivated by rewriting the integral representation as \eqref{g0:Aform} (see Appendix \ref{app:Aform}), completing \eqref{g0:closed}.

The closed-form \eqref{g0:closed} has been verified against the numerical Bethe potential data with great precision. First, we confirm that subtracting the exact expressions order by order from the numerical Bethe potential reduces the standard errors of the subleading fitting coefficients in \texttt{LinearModelFit} consecutively, which supports the closed-form \eqref{g0:closed}. Furthermore, the differences between the closed-form \eqref{g0:closed} and the numerical Bethe potential are at the level of $10^{-19}\sim 10^{-20}$ at all data points, precisely the level expected from the non-perturbative corrections
\begin{align}
	\hat g_{\rm np}~\sim~ e^{-2\pi\sqrt{2\lambda}}~\overset{\lambda=30}{\sim}~10^{-21}\,,
\end{align}
cf.\ \eqref{fnp}, which strongly implies that \eqref{g0:closed} captures the full perturbative constant. In particular, at $k=1$ it reproduces the value $\hat g_0=-0.18384102333840097854$ of Appendix~C.2 of \cite{Bobev:2022wem} in all 20 digits. We leave a first-principles derivation of the closed-form \eqref{g0:closed} for future work.

Remarkably, the closed-form \eqref{g0:closed} can be rewritten exactly as a finite linear combination of $A$-functions \eqref{Afct} as
\begin{align}
	\hat g_0(k,\Dsc)&=A(k)-A\Big(\fft k2\Big)+\fft{k}{2}A\Big(\fft4k\Big)-\fft{k}{4}A\Big(\fft8k\Big)\nn\\
	&\quad-\fft{\zeta(3)}{8\pi^2}k^2\,,\label{g0:Aform}
\end{align}
see Appendix~\ref{app:Aform} for the derivation. Combined with the exact values \eqref{Afct:special} of the $A$-function at integer arguments, the identity \eqref{g0:Aform} yields exact special values such as
\begin{align}
	\hat g_0(2,\Dsc)=-\fft{5\zeta(3)}{4\pi^2}\,,\qquad
	\hat g_0(4,\Dsc)=-\fft{3\zeta(3)}{2\pi^2}\,,\label{g0:special}
\end{align}
which precisely reproduce the numerical values $-0.15224228529196635390$ \& $-0.18269074235035962450$ at $k=2,4$ quoted in Appendix~C.2 of \cite{Bobev:2022wem}.

\medskip

\emph{Generic flavor chemical potentials.} We next turn to generic $\boldDelta$, using the numerical BAE data at $\lambda=30$ for the flavor configurations used in Appendix~D of \cite{Bobev:2022eus} and Appendix~C.2 of \cite{Bobev:2022wem}. The $k^2$ coefficient is known in closed-form as \eqref{g0:k2}. Our new result is the $k^0$ term of the large-$k$ expansion:
\begin{align}
	\hat g_0(k,\boldDelta)\Big|_{k^0}=-\fft{1}{24}\sum_{(a,b)\in\mathcal D}\Delta_{ab}\log\fft{2}{\Delta_{ab}}\,,\label{g0:const}
\end{align}
where the sum runs over the mixed pairs
\begin{align}
	\mathcal D=\{(1,3),(1,4),(2,3),(2,4)\}\,,\label{mixed:pairs}
\end{align}
distinguished by the interaction structure of the Bethe potential \eqref{TTI:V}; note that a full permutation symmetry is \emph{not} respected as in the $k^2$ coefficient. The compact expression \eqref{g0:const} is verified for all available flavor configurations within the typical numerical precision of \texttt{LinearModelFit}; \emph{e.g.} for $\boldDelta=(\fft{15}{40},\fft{17}{40},\fft{21}{40},\fft{27}{40})$, we find
\begin{align}
	\hat g_0^{\rm num}(k,\boldDelta)\Big|_{k^0}&=-0.115002956078388\,,\\
	\hat g_0(k,\boldDelta)\Big|_{k^0}&=-0.115002956076556\,.
\end{align}
The general expression \eqref{g0:const} also reduces to $-\fft16\log2$ at the superconformal point, consistent with \eqref{g0:closed}. Higher orders of the large-$k$ expansion at generic $\boldDelta$, and in particular the generalization of the full closed-form \eqref{g0:closed}, are left for future work.

\subsubsection{The topologically twisted index constant}\label{sec:ABJM:f0}

\emph{Superconformal point.} Our main result for the TTI constant at the superconformal point is the closed-form expression
\begin{widetext}
\begin{align}
	\hat f_0(k,\Dsc,\boldmn_{\rm sc})&=-\fft{3\zeta(3)}{8\pi^2}k^2+\fft16\log k+\mathfrak f_0
	+\fft{k}{\pi^2}\int_0^\infty dx\,\log\big(1-e^{-kx}\big)\bigg[12x\coth2x-2x\coth x-4+4\log\fft{\sinh x}{x}\bigg]\nn\\
	&=-\fft{3\zeta(3)}{8\pi^2}k^2+\fft16\log k+\mathfrak f_0
	+\sum_{n=1}^{\infty}\bigg(\fft{2\pi}{k}\bigg)^{2n}\fft{(-1)^n4^{n+1}\big(3n\cdot4^n-n+1\big)|B_{2n}B_{2n+2}|}{n\,(2n+2)!}\,,\label{f0:closed}
\end{align}
\end{widetext}
where the second equality is again the large-$k$ expansion of the integral representation (Appendix~\ref{app:series}), and the transcendental constant $\mathfrak f_0$ of \eqref{f0:series} is determined as
\begin{align}
	\mathfrak f_0=-8\zeta'(-1)-\fft{5}{2}\log2-\fft23\log4\pi\,,\label{frakf0}
\end{align}
which agrees with the numerical value quoted below \eqref{f0:series} in 17 digits.

The derivation logic parallels the Bethe potential case. The tail coefficients extracted from the TTI data are again exact rationals, from which --- motivated once more by the Bernoulli-number structure of \eqref{Afct} --- we deduced the general-order formula in the second line of \eqref{f0:closed}, i.e.\ the closed-form of the coefficients in \eqref{f0:series},
\begin{align}
	\mathfrak f_n=\fft{(-1)^n4^{n+1}\big(3n\cdot4^n-n+1\big)|B_{2n}B_{2n+2}|}{n\,(2n+2)!}\,.\label{f2n:formula}
\end{align}
It reproduces all five known fractions \eqref{f2n:known} and predicts the exact higher-order coefficients
\begin{align}
	\mathfrak f_6=\fft{3918661}{5746615875}\,,\qquad
	\mathfrak f_7=-\fft{1450417}{1138606875}\,,\label{f12f14}
\end{align}
and so on, which agree with an independent higher-order fit of the same data. The series is then resummed backward into the integral representation in the first line of \eqref{f0:closed} exactly as for the Bethe potential (Appendix~\ref{app:series}), with the integrand normalized to vanish at $x=0$ so that the split between the integral and $\mathfrak f_0$ is unambiguous. Recasting the resulting integral representation into the $A$-function combination (Appendix~\ref{app:Aform}) then exposes the transcendental basis $\{\zeta'(-1),\log2,\log4\pi\}$ of the constant channel; identifying the fitted $\mathfrak f_0$ within this basis --- in parallel with the identification of $-\fft{\log2}{6}$ in \eqref{g0:closed} --- yields \eqref{frakf0}, after which the compact expression \eqref{f0:Aform} is confirmed independently through the exact values \eqref{f0:special} below. The closed-form \eqref{f0:closed} has also been verified point-wise against the TTI data at $\lambda=30$, with agreement at the level of $10^{-18}$--$10^{-17}$; the residuals exceeding the bare exponential $e^{-2\pi\sqrt{2\lambda}}\sim10^{-21}$ of \eqref{fnp} can be understood via its $\mO(\log N)$ exponent, i.e.\ a polynomial prefactor whose detailed characterization from the quantum M2-brane path integral can be found in \cite{Gautason:2025per,vanMuiden:2026nsp}. 

As for the Bethe potential, the integral representation \eqref{f0:closed} can be recast into a finite linear combination of constant map functions (see Appendix~\ref{app:Aform}):
\begin{align}
	\hat f_0(k,\Dsc,\boldmn_{\rm sc})&=-4A(k)-3kA\Big(\fft4k\Big)+\fft{3k}{2}A\Big(\fft8k\Big)\nn\\
	&\quad+\fft{\zeta(3)}{4\pi^2}k^2-\fft12\log k-\fft52\log2\nn\\
	&=2A(k)-6A\Big(\fft k2\Big)-6\,\hat g_0(k,\Dsc)\nn\\
	&\quad-\fft{\zeta(3)}{2\pi^2}k^2-\fft12\log k-\fft52\log2\,,\label{f0:Aform}
\end{align}
where the second expression exhibits a direct linear relation between the two ABJM constants obtained in this paper. Since \eqref{f0:Aform} involves $A$ only at the arguments $\{k,\fft4k,\fft8k\}$, all integers for $k\in\{1,2,4\}$, the exact values \eqref{Afct:special} yield
\begin{align}
	\hat f_0(1,\Dsc,\boldmn_{\rm sc})&=\fft{21\zeta(3)}{16\pi^2}-\fft{37}{8}\log2\,,\nn\\
	\hat f_0(2,\Dsc,\boldmn_{\rm sc})&=\fft{21\zeta(3)}{4\pi^2}-\fft92\log2\,,\label{f0:special}
\end{align}
which reproduce the 20-digit numerical values $-3.0459513105331823845$ at $k=1$ and $-1.7865975337335498966$ at $k=2$ quoted in Appendix~C of \cite{Bobev:2022eus} precisely up to an additive constant $\log k$ from the degeneracy split discussed below \eqref{TTI:allorder}. 

\medskip
\noindent\emph{Note added.} The closed forms \eqref{f0:closed} and \eqref{f0:Aform} were independently obtained in \cite{Hosseini:2026companion}.

\medskip

\emph{Generic flavor chemical potentials.} The analogue of \eqref{g0:const} for the TTI constant $\hat f_0(k,\boldDelta,\boldmn)$ is not yet available: unlike the Bethe potential constant, it depends on the flavor magnetic fluxes as well as on the chemical potentials, and exploring this enlarged parameter space with sufficient numerical precision --- while straightforward in principle --- is substantially more laborious. We defer this analysis to a separate investigation and comment briefly on this point in Section~\ref{sec:discussion}. 

\section{Constants of the ADHM theory}\label{sec:ADHM}

The technology of Section~\ref{sec:ABJM} is not tied to the ABJM theory. In this section we apply it to the $\mathrm{U}(N)$ $\mathcal N=4$ ADHM theory, whose Bethe potential and TTI constants have likewise been available only numerically \cite{Bobev:2022wem,Bobev:2023lkx}. In parallel with the previous section, we first briefly introduce the theory, specialize the localization formulae of Section~\ref{sec:setup} to it, and review the known all-order results. We then determine the Bethe potential constant in closed-form and present the first few exact coefficients in the large-$\Nf$ expansion of the TTI constant.

\subsection{Review}\label{sec:ADHM:review}
The ADHM theory is the 3d $\mathcal N=4$ $\mathrm{U}(N)$ gauge theory with vanishing CS level, one adjoint hypermultiplet and $\Nf$ fundamental hypermultiplets, describing the low-energy dynamics of $N$ M2-branes probing a $\mathbb C^2\times(\mathbb C^2/\mathbb Z_{\Nf})$ singularity, holographically dual to M-theory on AdS$_4\times S^7/\mathbb Z_{\Nf}$ with a fixed point \cite{Benini:2009qs,Bashkirov:2010kz,Mezei:2013gqa,Grassi:2014vwa}; the name refers to its role in the ADHM construction of instantons \cite{Atiyah:1978ri}, and the theory is also known as the $\Nf$ model \cite{Grassi:2014vwa,Minahan:2021pfv}. At $\Nf=1$ the ADHM theory is dual to the ABJM theory at $k=1$ under 3d mirror symmetry \cite{Aharony:2008ug,Kapustin:2010xq}. In $\mathcal N=2$ language it contains three adjoint chiral multiplets $\Psi_I$ $(I=1,2,3)$ and $\Nf$ pairs of fundamental and anti-fundamental chiral multiplets $\psi_q$ and $\tpsi_q$ $(q=1,\ldots,\Nf)$, interacting through the superpotential 
\begin{align}
	W=\Tr\bigg[\sum_{q=1}^{\Nf}\tpsi_q\Psi_3\psi_q+\Psi_3[\Psi_1,\Psi_2]\bigg]\,.\label{ADHM:W}
\end{align}
In addition to the $\mathrm{SO}(4)$ R-symmetry, the theory enjoys an $\mathrm{SU}(2)\times\mathrm{SU}(\Nf)$ flavor symmetry as well as a topological $\mathrm{U}(1)$ symmetry. 

In the localization formulae of the ADHM TTI, the chemical potentials $\Delta_I$ and $(\Delta_q,\Delta_{\tilde q})$ of the adjoint and (anti-)fundamental chiral multiplets, the associated fluxes $(\mn_I,\mn_q,\mn_{\tilde q})$, and the chemical potential $\Delta_m$ and background flux $\mkt$ of the topological $\mathrm{U}(1)$ appear, subject to the constraints following from the marginality of the superpotential \eqref{ADHM:W} 
\begin{equation}
\begin{alignedat}{2}
	\Delta_1+\Delta_2+\Delta_3&=\Delta_q+\Delta_{\tilde q}+\Delta_3&&=2\,,\\
	\mn_1+\mn_2+\mn_3&=\mn_q+\mn_{\tilde q}+\mn_3&&=2\,.\label{ADHM:constraints}
\end{alignedat}
\end{equation}
For the ADHM theory, the BA formula \eqref{BA:general} takes the explicit form 
\begin{widetext}
\begin{align}
	Z^{\rm ADHM}_{S^1\times S^2}(N,\Nf,\boldDelta,\boldmn)
	&=y_q^{\fft{N\Nf(1-\mn_q)}{2}}y_{\tilde q}^{\fft{N\Nf(1-\mn_{\tilde q})}{2}}\prod_{I=1}^3y_I^{-\fft{N^2}{2}\mn_I}
	\sum_{\{x_i\}\in\mathrm{BAE}}\fft{1}{\det\mathbb B}\,
	\fft{\prod_{i=1}^Nx_i^{N+\mkt}\prod_{i\neq j}^N\big(1-\fft{x_i}{x_j}\big)}{\prod_{I=1}^3\prod_{i,j=1}^N(x_j-x_iy_I)^{\fft{1-\mn_I}{2}}(x_i-x_jy_I)^{\fft{1-\mn_I}{2}}}\nn\\
	&\quad\times\prod_{i=1}^N\fft{x_i^{\Nf(1-\fft{\mn_q+\mn_{\tilde q}}{2})}}{(1-x_iy_q)^{\Nf(1-\mn_q)}(x_i-y_{\tilde q})^{\Nf(1-\mn_{\tilde q})}}\,,\label{ADHM:TTI:BA}
\end{align}
with $x_i=e^{\ri u_i}$ the $\mathrm{U}(N)$ holonomies, $y=e^{\ri\pi\Delta}$ for each flavor label, and $\mathbb B$ the BAE Jacobian; see \cite{Bobev:2023lkx} for more details. The ADHM Bethe potential reads 
\begin{align}
	\mV^{\rm ADHM}(N,\Nf,u,\boldDelta)&=\sum_{i=1}^N\bigg[\Big(2n_i+N+1-2\Big\lfloor\fft{N+1}{2}\Big\rfloor-\Delta_m\Big)\pi u_i-\fft{\Nf\pi}{2}\big(2-\Delta_q-\Delta_{\tilde q}\big)u_i\bigg]\nn\\
	&\quad+\fft12\sum_{I=1}^3\sum_{i,j=1}^N\Big[\mathrm{Li}_2\big(e^{\ri(u_j-u_i+\pi\Delta_I)}\big)-\mathrm{Li}_2\big(e^{\ri(u_j-u_i-\pi\Delta_I)}\big)\Big]\nn\\
	&\quad+\Nf\sum_{i=1}^N\Big[\mathrm{Li}_2\big(e^{\ri(-u_i+\pi\Delta_{\tilde q})}\big)-\mathrm{Li}_2\big(e^{\ri(-u_i-\pi\Delta_q)}\big)\Big]\,,\label{ADHM:V}
\end{align}
\end{widetext}
with the integers chosen as $n_i=\lfloor\fft{N+1}{2}\rfloor-i$ for the standard large-$N$ BAE solution. Following \cite{Bobev:2022wem,Bobev:2023lkx,Bobev:2025ltz,Bobev:2026lvl} we trade the flavor parameters for the ``tilde'' variables 
\begin{align}
	\boldtDelta&=\Big(\Delta_1,\,\Delta_2,\,\fft{\Delta_3}{2}-\fft{\Delta_m}{\Nf},\,\fft{\Delta_3}{2}+\fft{\Delta_m}{\Nf}\Big)\,,\nn\\
	\boldtmn&=\Big(\mn_1,\,\mn_2,\,\fft{\mn_3}{2}+\fft{\mkt}{\Nf},\,\fft{\mn_3}{2}-\fft{\mkt}{\Nf}\Big)\,,\label{ADHM:tilde}
\end{align}
in terms of which the ADHM constraints \eqref{ADHM:constraints} take the same forms as the ABJM ones \eqref{ABJM:constraints}. The superconformal and universal configuration is 
\begin{align}
	\tDsc:~\tD_a=\fft12\,,\qquad \boldtmn_{\rm sc}:~\tmn_a=\fft12\,.\label{ADHM:sc:config}
\end{align}

\medskip

The imaginary part of the ADHM Bethe potential evaluated on the exact numerical BAE solution is shown to take the all-order form \cite{Bobev:2022wem,Bobev:2023lkx} 
\begin{align}
	\fft{\Im\mV^{\rm ADHM}(N,\Nf,\boldtDelta)}{2\pi}&=\fft{\pi\sqrt{2\Nf\tD_1\tD_2\tD_3\tD_4}}{3}\,\hat N_{\Nf,\boldtDelta}^{3/2}\nn\\
	&\hspace{-6em}+\hat g_0^{\rm ADHM}(\Nf,\boldtDelta)+\hat g^{\rm ADHM}_{\rm np}(N,\Nf,\boldtDelta)\,,\label{ADHM:ImV}
\end{align}
where the shifted rank is given by \cite{Bobev:2023lkx} 
\begin{align}
	\hat N_{\Nf,\boldtDelta}\equiv N-\fft{\Nf}{24}+\fft{\Nf}{12}\big(\tD_1^{-1}+\tD_2^{-1}\big)+\fft{\tD_3^{-1}+\tD_4^{-1}}{12\Nf}\,.\label{ADHM:Nhat}
\end{align}
The all-order $1/N$ expansion of the ADHM TTI similarly reads \cite{Bobev:2022wem,Bobev:2023lkx} 
\begin{widetext}
\begin{align}
	\log Z^{\rm ADHM}_{S^1\times S^2}(N,\Nf,\boldtDelta,\boldtmn)&=-\fft{\pi\sqrt{2\Nf\tD_1\tD_2\tD_3\tD_4}}{3}\sum_{a=1}^4\tmn_a\bigg[\fft{1}{\tD_a}\hat N_{\Nf,\boldtDelta}^{3/2}+\Big(\mathfrak c_a(\boldtDelta)\Nf+\fft{\mathfrak d_a(\boldtDelta)}{\Nf}\Big)\hat N_{\Nf,\boldtDelta}^{1/2}\bigg]\nn\\
	&\quad-\fft12\log\hat N_{\Nf,\boldtDelta}+\hat f_0^{\rm ADHM}(\Nf,\boldtDelta,\boldtmn)+\hat f^{\rm ADHM}_{\rm np}(N,\Nf,\boldtDelta,\boldtmn)\label{ADHM:TTI:allorder}
\end{align}
\end{widetext}
up to an overall phase, with the rational functions $\mathfrak c_a,\mathfrak d_a$ given in \cite{Bobev:2023lkx} and the non-perturbative corrections denoted $\hat f^{\rm ADHM}_{\rm np}$. The constants $\hat g_0^{\rm ADHM}$ and $\hat f_0^{\rm ADHM}$ were determined numerically for various $\Nf$ and flavor configurations \cite{Bobev:2022wem,Bobev:2023lkx,Bobev:2025ltz}, but no closed-forms have been available so far.

\subsection{New results}\label{sec:ADHM:new}

The numerical analysis toward the closed-form expression for the $N$-independent constants $\hat g_0^{\rm ADHM}$ and $\hat f_0^{\rm ADHM}$ is parallel to that of Section~\ref{sec:ABJM:new} for the ABJM theory. We first solve the ADHM BAE $\partial_{u_i}\mV^{\rm ADHM}=0$ numerically as in \cite{Bobev:2023lkx} at the superconformal configuration \eqref{ADHM:sc:config} and fixed 't~Hooft coupling $\lambda\equiv N/\Nf$. For the Bethe potential we use $\lambda=30$ with $N=11\sim401$ in steps of $10$, while for the TTI we generate two independent data sets $\lambda\in\{20,30\}$, each with $N=11\sim551$ ($501$ for $\lambda=30$) in steps of $10$. Generic flavor configurations are left for future work.

\subsubsection{The Bethe potential constant}\label{sec:ADHM:g0}

Our main result for the ADHM Bethe potential constant is the closed-form expression
\begin{widetext}
\begin{align}
	\hat g_0^{\rm ADHM}(\Nf,\tDsc)&=\mathsf a_2\,\Nf^2-\fft{\log2}{8}
	+\fft{\Nf}{\pi^2}\int_0^\infty dx\,\log\big(1-e^{-\Nf x}\big)\bigg[\fft12\log\cosh x+\fft14\log\cosh\fft x2\bigg]\nn\\
	&=\mathsf a_2\,\Nf^2-\fft{\log2}{8}+\sum_{n=1}^{\infty}\bigg(\fft{2\pi}{\Nf}\bigg)^{2n}\fft{(-1)^n(4^n-1)(2\cdot 4^n+1)|B_{2n}B_{2n+2}|}{4n\,(2n+2)!}\label{ADHMg0:closed}
\end{align}
\end{widetext}
with the leading order coefficient
\begin{align}
	\mathsf a_2=\fft{3\zeta(3)}{32\pi^2}+\fft{7}{32}\log2-\fft12A\Big(\fft12\Big)\,.\label{ADHMg0:a2}
\end{align}
The expression \eqref{ADHMg0:closed} was obtained by the three-step procedure of Section~\ref{sec:ABJM:g0} applied verbatim to the ADHM data. The only structural difference from \eqref{g0:closed} is the two-argument combination $\fft12\log\cosh x+\fft14\log\cosh\fft x2$ in the integrand, from which we derive the exact two-level relation between the ADHM and ABJM Bethe potential constants
\begin{align}
	\hat g_0^{\rm ADHM}(\Nf,\tDsc)&=\fft12\hat g_0(\Nf)+\fft14\hat g_0(2\Nf)\nn\\
	&\quad+\bigg(\fft12\hat g_0(1)-\fft14\hat g_0(2)\bigg)\Nf^2\,, \label{ADHMg0:master}
\end{align}
where $\hat g_0(k)\equiv\hat g_0(k,\Dsc)$ denotes the ABJM constant \eqref{g0:closed}. Note that \eqref{ADHMg0:master} manifestly respects the 3d mirror symmetry: at $\Nf=1$ the ADHM theory is dual to the ABJM theory at $k=1$, so $\hat g_0^{\rm ADHM}(1,\tDsc)=\hat g_0(1)$. In fact, this mirror duality constraint was the key in determining the leading coefficient \eqref{ADHMg0:a2} which involves a highly non-trivial transcendental number $A(1/2)=0.44522628637808630409\ldots$ that lies outside the integer arguments for which the $A$-function is known in elementary closed-form \cite{Hatsuda:2014vsa}.

\medskip

The closed-form \eqref{ADHMg0:closed} has been verified against the numerical Bethe potential data with great precision, in parallel with the ABJM case. First, subtracting the exact expressions order by order from the numerical Bethe potential consecutively reduces the standard errors of the subleading fitting coefficients in \texttt{LinearModelFit}, supporting the closed-form \eqref{ADHMg0:closed}. Moreover, the point-wise differences between \eqref{ADHMg0:closed} and the numerical Bethe potential are at the level expected from the non-perturbative corrections $\hat g^{\rm ADHM}_{\rm np}\sim e^{-2\pi\sqrt{2\lambda}}\sim10^{-21}$ at $\lambda=30$ \cite{Bobev:2023lkx}. 

As in the ABJM case, the closed-form \eqref{ADHMg0:closed} can also be rewritten exactly as a finite linear combination of $A$-functions: inserting \eqref{g0:Aform} into \eqref{ADHMg0:master} gives
\begin{align}
	\hat g_0^{\rm ADHM}(\Nf,\tDsc)&=\fft14A(\Nf)-\fft12A\Big(\fft{\Nf}{2}\Big)+\fft14A(2\Nf)\nn\\
	&\quad+\fft{\Nf}{4}A\Big(\fft{2}{\Nf}\Big)+\fft{\Nf}{8}A\Big(\fft4{\Nf}\Big)\nn\\
	&\quad-\fft{\Nf}{8}A\Big(\fft8{\Nf}\Big)+\bigg(\mathsf a_2-\fft{9\zeta(3)}{64\pi^2}\bigg)\Nf^2\,,\label{ADHMg0:Aform}
\end{align}
exhibiting the same characteristic pattern of direct arguments $\{\Nf,\fft{\Nf}{2},2\Nf\}$ and inverted arguments $\{\fft{2}{\Nf},\fft{4}{\Nf},\fft{8}{\Nf}\}$ as in \eqref{g0:Aform}. 

\subsubsection{The topologically twisted index constant}\label{sec:ADHM:f0}

Applying the same analysis to the ADHM TTI data and extracting the $N$-independent constant through \eqref{ADHM:TTI:allorder}, we find that $\hat f_0^{\rm ADHM}$ at the superconformal configuration \eqref{ADHM:sc:config} organizes as
\begin{align}
	\hat f_0^{\rm ADHM}(\Nf,\tDsc,\boldtmn_{\rm sc})&=\mathsf k_2\,\Nf^2+\fft13\log\Nf+\mathsf k_0\nn\\
	&\quad+\sum_{n=1}^{\infty}\bigg(\fft{2\pi}{\Nf}\bigg)^{2n}\,\mathsf t_n\,,\label{ADHMf0:data}
\end{align}
with (quoted from the $\lambda=30$ data)
\begin{align}
	\mathsf k_2&=-0.159517636940496\,,\nn\\
	\mathsf k_0&=-2.211153205899\,.\label{ADHMf0:numbers}
\end{align}
Notably, the $\log\Nf$ coefficient left free in the fit returns $0.3333333333334$ with the standard error $2.082\times 10^{-15}$ --- \emph{twice} the ABJM value $\fft16$ of \eqref{f0:series}.

The first four tail coefficients are determined as exact rationals as
\begin{align}
	\{\mathsf t_n\}_{n\leq4}=\Big\{&-\fft{41}{1440},\,\fft{103}{72576},\nn\\
	&-\fft{1429}{4354560},\,\fft{103847}{638668800}\Big\}\,,\label{ADHMf0:tail}
\end{align}
with each identified rational reproducing the fitted value with great precision, further improving the subleading fits upon its analytic subtraction. The closed-form resummation of the large-$\Nf$ expansion \eqref{ADHMf0:data}, in analogy with \eqref{f0:closed} --- for which the exact mirror symmetry value
\begin{align}
	\hat f_0^{\rm ADHM}(1,\tDsc,\boldtmn_{\rm sc})&=\hat f_0(1,\Dsc,\boldmn_{\rm sc})\nn\\
	&=\fft{21\zeta(3)}{16\pi^2}-\fft{37}{8}\log2\label{ADHMf0:mirror}
\end{align}
from \eqref{f0:special} provides an exact anchor --- is left for future work.

\section{Beyond the Bethe potential and TTI: the Airy constant}\label{sec:Airy}

Sections~\ref{sec:ABJM} and \ref{sec:ADHM} were devoted exclusively to the Bethe potential and the TTI. We now turn to the squashed sphere partition function of Section~\ref{sec:setup} and show how those closed-forms determine its Airy constant $\mA$ in \eqref{Airy} through the two leading orders of its large-squashing expansion --- either small or large $b$, the two being related by the squashing-inversion symmetry $b\to b^{-1}$ --- using the factorization relations recently established in \cite{Bobev:2026lvl}. We then recall the exact values of $\mA$ for special cases that are known in the literature, collect them with our new results into the web of independent anchors that any candidate closed-form for $\mA$ at generic squashing must satisfy, and comment on the remaining steps toward a complete determination. 

\subsection{Large-squashing expansion of the Airy constant}\label{sec:Airy:fact}

Throughout this section we follow \cite{Bobev:2026lvl} and trade the squashing parameter $b$ and the flavor chemical potentials for 
\begin{align}
	\omega\equiv b^2\,,\quad \varphi_a\equiv(1+\omega)\Delta_a\,,\label{omega:varphi}
\end{align}
with $\Delta_a\to\tD_a$ understood for the ADHM theory. In these variables the Airy coefficient $\mC$ of \eqref{Airy:BC} takes the compact form 
\begin{align}
	\mC(\omega,K,\boldvarphi)=\fft{2\omega^2}{\pi^2K\,\varphi_1\varphi_2\varphi_3\varphi_4}\label{Airy:C}
\end{align}
%
for both theories of interest with 
\begin{align}
	K=\begin{cases}
		k & (\text{ABJM}) \\
		\Nf & (\text{ADHM})
	\end{cases}\,.
\end{align}
The coefficient $\mB$, which we shall not need, differs between the two theories and is recorded in  \cite{Bobev:2025ltz,Bobev:2026lvl}.

It was shown in \cite{Bobev:2026lvl} that both the SCI and the TTI of the class of theories under consideration factorize into pairs of squashed sphere partition functions to all orders in the Cardy-like expansion. Focusing on the $N$-independent sector, the two Airy factors arising in the factorized SCI share the \emph{same} squashing $\omega$, while those of the TTI carry \emph{opposite} squashings $\pm\omega$; at the superconformal configuration \eqref{sc:config} such Airy factors sit on the two slices
\begin{align}
	\mA(\pm\omega,K)\equiv\mA\big(\pm\omega,K,\boldvarphi^\pm\big)\,,\quad \varphi^\pm_a=\fft{1\pm\omega}{2}\,.\label{mA:shorthand}
\end{align}
Taking the $N$-independent part of the two factorization relations then gives, for the SCI and the (unrefined) TTI respectively,
\begin{align}
	\log\mI\,\big|_{N\text{-indept}}&=2\mA(\omega,K)-\fft12\log\mC\big(\omega,K,\boldvarphi^+\big)\nn\\
	&\quad-\log4\pi\,,\label{fact:SCI}\\
	\log Z_{S^1\times S^2}\big|_{N\text{-indept}}&-\log k=\mA(\omega,K)+\mA(-\omega,K)\nn\\
	&\quad+\log\fft{\mC^{-1/4}\,\widetilde{\mC}^{-1/4}}{4\pi}+\mO(\omega)\,,\label{fact:TTI}
\end{align}
where $\mC=\mC(\omega,K,\boldvarphi^+)$ and $\widetilde{\mC}=\mC(-\omega,K,\boldvarphi^-)$, and the explicit $-\log k$ implements the \emph{single} Bethe vacuum isolated by the factorization [absent for ADHM]. We refer the reader to Sections~4 and 5 of \cite{Bobev:2026lvl} for the detailed derivation of \eqref{fact:SCI} and \eqref{fact:TTI}, with one caveat --- the factorization conventions are not precisely identical to those used frequently in the analysis of indices and therefore the dictionary must be applied with care. 

The left-hand sides of \eqref{fact:SCI} and \eqref{fact:TTI} are precisely the quantities computed in Sections~\ref{sec:ABJM} and \ref{sec:ADHM}: the Cardy-like expansion \eqref{SCI:Cardy} gives $\log\mI\,|_{N\text{-indept}}=-\fft{2}{\omega}\hat g_0+\hat f_0+\mO(\omega)$ by \eqref{ImV} and \eqref{TTI:allorder}, while $\log Z_{S^1\times S^2}|_{N\text{-indept}}-\log k=\hat f_0$. Combining these with the factorization relations \eqref{fact:SCI} and \eqref{fact:TTI} yields the following results.
\begin{itemize}
	\item Inserting the former into \eqref{fact:SCI} with $\mC(\omega,K,\boldvarphi^+)=\fft{32\,\omega^2}{\pi^2K(1+\omega)^4}$ from \eqref{Airy:C} and expanding $\log\mC$ at small $\omega$ yields 
	\begin{align}
		\mA(\omega,K)&=-\fft{1}{\omega}\,\hat g_0+\fft14\log\omega^2 \label{mA:smallb}\\
		&\quad+\fft12\hat f_0+\fft94\log2-\fft14\log K+\mO(\omega)\nn
	\end{align}
	where $\hat g_0$ and $\hat f_0$ are the superconformal Bethe potential and the TTI constants of the corresponding theory. Consequently, the Airy constant of the squashed sphere partition function is determined \emph{in closed-form} through the two leading orders of its large-squashing expansion by the constants computed in this paper, while the structural aspect of \eqref{mA:smallb} has already been put forward in \cite{Bobev:2022wem,Bobev:2025ltz,Bobev:2026lvl}.
	
	\item Likewise, inserting $\log Z_{S^1\times S^2}|_{N\text{-indept}}-\log k=\hat f_0$ into \eqref{fact:TTI} with $\mC^{-1/4}\widetilde{\mC}^{-1/4}=\fft{\pi K^{1/2}}{\sqrt{32}}\,(\omega^2)^{-1/2}(1-\omega^2)$ from \eqref{Airy:C} fixes the part of the Airy constant that is even under $\omega\to-\omega$ as
	\begin{align}
		\mA(\omega,K)+\mA(-\omega,K)&=\hat f_0+\fft12\log\omega^2-\fft12\log K\nn\\
		&\quad+\fft92\log2+\mO(\omega)\,,\label{mA:even}
	\end{align}
	which follows from \eqref{mA:smallb} automatically.
\end{itemize}
%

\medskip

For the ABJM theory, all functions of $k$ on the right-hand side of \eqref{mA:smallb} are now known exactly. Substituting the closed-forms \eqref{g0:Aform} and \eqref{f0:Aform} into \eqref{mA:smallb} yields the large-squashing expansion of the superconformal Airy constant for the ABJM theory as
\begin{align}
	\mA(\omega,k)&=-\fft{1}{\omega}\,\hat g_0(k)+\fft14\log\omega^2\nn\\
	&\quad-2A(k)-\fft{3k}{2}A\Big(\fft4k\Big)+\fft{3k}{4}A\Big(\fft8k\Big)\nn\\
	&\quad+\fft{\zeta(3)}{8\pi^2}k^2-\fft12\log k+\log2+\mO(\omega)\,.\label{mA:smallb:ABJM}
\end{align}
The large-squashing expansion \eqref{mA:smallb:ABJM} passes non-trivial consistency checks against previously known data in the large $k$ limit via \eqref{Afct}: the $k^2$ coefficient of \eqref{mA:smallb:ABJM} reproduces the exact planar result ($Q=b+b^{-1}$) \cite{Bobev:2025ltz} 
\begin{align}
	\mA(\omega,k)\Big|_{k^2}=\fft{\zeta(3)}{8\pi^2}\Big(\fft{Q^2}{4}-2\Big)k^2 \label{mA:k2}
\end{align}
at orders $\omega^{-1}$ and $\omega^0$, and the $\log k$ coefficient in \eqref{mA:smallb:ABJM} equals $-\fft16$, in precise agreement with the universal value conjectured in \cite{Bobev:2026lvl} even beyond $\mO(\omega)$.

\medskip

For the ADHM theory, specializing \eqref{mA:smallb} gives
\begin{align}
	\mA^{\rm ADHM}(\omega,\Nf)&=-\fft{1}{\omega}\,\hat g_0^{\rm ADHM}(\Nf,\tDsc)+\fft14\log\omega^2\nn\\
	&\quad+\fft12\hat f_0^{\rm ADHM}(\Nf,\tDsc,\boldtmn_{\rm sc})\nn\\
	&\quad+\fft94\log2-\fft14\log\Nf+\mO(\omega)\,,\label{ADHM:mA:smallb}
\end{align}
exact through the two leading orders of the large-squashing expansion. Here the leading $\omega^{-1}$ coefficient is known in closed-form via \eqref{ADHMg0:closed}, whereas the constant $\hat f_0^{\rm ADHM}$ entering the $\mO(\omega^0)$ order is presently characterized by the numerical data \eqref{ADHMf0:numbers} and the exact tail coefficients \eqref{ADHMf0:tail} rather than by a closed-form. We can still extract an interesting observation from \eqref{ADHM:mA:smallb}: it determines the $\log\Nf$ coefficient in the large $\Nf$ expansion of the ADHM Airy constant in principle up to $\mO(\omega)$ but possibly beyond, for which no result at generic squashing is available in the literature. Specifically, the $\fft13\log\Nf$ term of \eqref{ADHMf0:data} combined with the explicit $-\fft14\log\Nf$ in \eqref{ADHM:mA:smallb} yields the $\log\Nf$ coefficient $-\fft1{12}$, which agrees at $\omega=1$ with the large-$\Nf$ expansion of the round sphere $A$-function expression $\mA^{\rm ADHM}(1,\Nf,\tDsc)=\fft12A(\Nf)+\fft12A(1)\,\Nf^2$, see \cite{Hatsuda:2014vsa,Bobev:2025ltz} for example. This consistency beyond the $\mO(\omega)$ correction, which is observed above in the ABJM case as well, strongly supports a universal $\log\Nf$ coefficient motivated in the dual 1-loop supergravity analysis \cite{Hristov:2021zai,Bobev:2023dwx}. 

\subsection{Toward the full Airy constant}\label{sec:Airy:anchors}
The results of Section~\ref{sec:Airy:fact} determine the two leading orders in the large-squashing expansion of the superconformal Airy constant mostly in closed-form. It is natural to ask for the full function $\mA(\omega,K)$ at generic squashing $\omega=b^2$ --- and, more ambitiously, its mass deformations. While we do not solve this problem here, the closed-forms of this paper substantially enlarge the web of exact constraints available to test any candidate expression, and we find it useful to collect the independent anchors in one place. Hence we present such anchors for the ABJM theory below.

\begin{itemize}
	\item[(A1)] \emph{Large-squashing expansion, all $k$}: the closed-form \eqref{mA:smallb:ABJM}, exact through $\mO(\omega^0)$ --- the main result of this paper.

	\item[(A2)] \emph{Round sphere, all $k$, two real masses}: the compact Fermi-gas expression of \cite{Nosaka:2015iiw} 
	\begin{align}
		\mA^{\rm ABJM}(1,k,\boldDelta_{\rm Nosaka})=\fft14\sum_{a=1}^4A(2\Delta_ak)\label{mA:Nosaka}
	\end{align}
	on the two-mass configuration $\boldDelta_{\rm Nosaka}=(\Delta_1,\Delta_2,1-\Delta_2,1-\Delta_1)$, reducing to $\mA(1,k)=A(k)$ at the superconformal point \cite{Marino:2011eh,Hanada:2012si,Hatsuda:2014vsa}.

	\item[(A3)] \emph{Squashing--mass relation, all $k$ and $b$}: the matrix model symmetry of \cite{Chester:2021gdw,Minahan:2021pfv}, valid at finite $N$ and $k$, implies for the Airy constant 
	\begin{align}
		&\mA\Big(b,k,m_1,m_2,\ri\,\fft{b-b^{-1}}{2}\Big)\label{mA:massmap}\\
		&=\mA\Big(1,k,\fft{b^{-1}m_++b\,m_-}{2},\fft{b^{-1}m_+-b\,m_-}{2},0\Big)\,,\nn
	\end{align}
	where $m_\pm\equiv m_1\pm m_2$ and the last three arguments denote the real masses, related to the flavor chemical potentials as $\Delta_{1,2}=\fft12-\ri\,\fft{m_1\pm(m_2+m_3)}{Q}$ and $\Delta_{3,4}=\fft12+\ri\,\fft{m_1\pm(m_2-m_3)}{Q}$. 

	\item[(A4)] \emph{Squashed Fermi gas at $k=1$}: for generic $b^2\in2\mathbb N-1$ and the two mass parameters of (A2) parametrized by $(m_1,m_2,m_3)=(\zeta-m,-\zeta-m,0)$ with one of them related to the squashing as $m=\fft{(b^2-3)\ri}{4b}$, $\mA$ is given by a four-term $A$-function combination derived in \cite{Kubo:2024qhq}. At $b^2=3$ it reproduces the Hatsuda point $\mA^{\rm ABJM}(\sqrt3,1,\Dsc)=\fft34A(2)-\fft14A(6)$ \cite{Hatsuda:2016uqa}. 

	\item[(A5)] \emph{Planar limit, all $\omega$}: the superconformal $k^2$ coefficient \eqref{mA:k2}, exact for all squashing; its generalization to arbitrary flavor chemical potentials is given in \cite{Bobev:2025ltz}.

	\item[(A6)] \emph{Universal $\log k$}: the large-$k$ coefficient of $\log k$ is conjectured to equal $-\fft16$ for all $\omega$ \cite{Bobev:2026lvl}; it is consistent with (A1) at large squashing but remains to be established rigorously at generic $\omega$.
\end{itemize}
A parallel list holds for the ADHM theory, whose entries can be found in \cite{Bobev:2025ltz} together with \eqref{ADHM:mA:smallb}. Notably, the special-squashing information is richer for the ADHM theory: the analogue of (A4) --- the value at generic $b^2\in2\mathbb N-1$ with the aligned mass configuration --- is available for \emph{all} $\Nf$ \cite{Bobev:2025ltz}, whereas the corresponding ABJM results are restricted to $k=1$ \cite{Kubo:2024qhq}. 

One further exact relation deserves to be recorded, since it follows from the consistency of the anchor web rather than adding to it. At $k=1$, tuning the free parameter of the (A4) configuration to $\zeta=-\fft{\ri(3b^2-5)}{4b}$ places it on the locus of (A3): the real masses of the (A4) configuration evaluate at this $\zeta$ to $(m_1,m_2,m_3)=\big({-}\ri\fft{b^2-2}{b},\,\ri\fft{b-b^{-1}}{2},\,0\big)$, which has the special third mass of \eqref{mA:massmap} upon the relabeling $m_2\leftrightarrow m_3$ (corresponding to the $\Delta_3\leftrightarrow\Delta_4$ symmetry). 
Evaluating $\mA$ there in two ways --- via \eqref{mA:massmap} followed by \eqref{mA:Nosaka}, and directly via (2.33) of \cite{Bobev:2025ltz} --- gives
\begin{align}
	\mA&=\fft14\Big[A\Big(\fft{2}{b^2}\Big)+A\Big(2-\fft{2}{b^2}\Big)\nn\\
	&\kern4em+A(3-b^2)+A(b^2-1)\Big]\nn\\
	&=\fft14\Big[A(2b^2-2)-A(2b^2)\nn\\
	&\kern4em+A(3-b^2)+A(b^2-1)\Big]\,,\label{mA:overlap}
\end{align}
whose equality, upon using the odd parity $A(-k)=-A(k)$, is precisely the functional identity
\begin{align}
	\Phi(x)=-\Phi\Big(\fft4x\Big)\,,\quad \Phi(x)\equiv A(x)+A(2-x)\,,\label{A:identity}
\end{align}
evaluated at $x=\fft{2}{b^2}$. Since (2.33) of \cite{Bobev:2025ltz} is established for $b^2\in2\mathbb N-1$, the anchor web thereby derives \eqref{A:identity} on the infinite discrete set $x\in\{\fft23,\fft25,\fft27,\ldots\}$; we have confirmed the full functional identity numerically to 40 significant digits over continuous ranges of $x$, and to the best of our knowledge it has not appeared in the literature. At $x=1$ it reduces to the special-value relation $2A(1)=A(2)-A(4)$ based on \eqref{Afct:special}. It would be very interesting to derive \eqref{A:identity} directly from the integral representation \eqref{Afct}, independently of the anchor web.

Presenting this web of exact constraints as a guide, we leave the complete determination of the full Airy constant $\mA(\omega,k,\boldvarphi)$ to future work.

\section{Discussion}\label{sec:discussion}

In this work we have determined, in closed-form, the $N$-independent constants governing the all-order $1/N$ expansions of the Bethe potential and of the TTI of the ABJM theory at the superconformal point, as well as the Bethe potential constant of the ADHM theory, and we have shown that all of them involve finite linear combinations of the constant map function $A$ evaluated at the direct and inverted arguments $\{K,\fft K2,2K;\fft2K,\fft4K,\fft8K\}$ with $K\in\{k,\Nf\}$. Through the factorization relations of \cite{Bobev:2026lvl} these results determine the Airy constant of the squashed sphere partition function to the two leading orders of its large-squashing expansion. We close with the open questions we find most pressing.
\begin{itemize}
	\item \emph{Generic flavor configurations.} The most immediate extension is to promote our superconformal closed-forms to generic flavor chemical potentials and fluxes, initiated here in \eqref{g0:const}. Derivatives of supersymmetric free energies with respect to real masses and squashing compute correlation functions of flavor currents and stress tensors integrated over the sphere \cite{Closset:2012vg,Closset:2012ru}, a link that has been developed into precision M-theory data  \cite{Agmon:2017xes,Chester:2018aca,Binder:2018yvd,Binder:2019mpb,Agmon:2019imm,Chester:2024bij}. Closed-form flavor dependence of the $N$-independent constants would thus pin down the constant map contribution to such integrated correlators --- precisely the piece not captured by the polynomial Airy data.
	
	\item \emph{Other M2-brane theories and beyond.} The strategy of Sections~\ref{sec:ABJM} and \ref{sec:ADHM} applies verbatim to the other classes with all-order TTI expansions such as the theories dual to AdS$_4\times N^{0,1,0}/\mathbb Z_k$, $V^{5,2}/\mathbb Z_{\Nf}$ and $Q^{1,1,1}/\mathbb Z_{\Nf}$ \cite{Bobev:2023lkx}, for which one may ask whether the constants are again finite $A$-function combinations. Beyond the M2-brane class, natural targets include the 3d $\mathcal N=2$ theories from wrapped M5-branes, whose perturbative $1/N$ expansions terminate at finite order \cite{Gang:2018hjd,Gang:2019uay,Bobev:2020zov}, and theories with massive type IIA duals \cite{Gaiotto:2009mv,Guarino:2015jca}, where the analogue of the constant term is essentially unexplored.
	
	\item \emph{Physical origin of the closed-forms.} On the round sphere the function $A(k)$ arises from resumming the constant map contributions of the topological string on the dual geometry \cite{Marino:2009jd,Drukker:2010nc,Hanada:2012si,Hatsuda:2012dt,Hatsuda:2014vsa}. No such description based on a duality chain to topological strings exists for the Bethe potential or the TTI, and yet their constants take a similar $A$-functional form. It would be very interesting to explore the hidden physics across these closed-form expressions, which will also shed light on the genus expansion of the TTI in the IIA regime as discussed in \cite{Bobev:2022jte,Bobev:2022eus}.
	
	\item \emph{Relation to gravity-side derivations.} The all-order Airy structure has recently been analyzed from several complementary bulk perspectives --- higher-derivative supergravity and gravitational blocks \cite{Hristov:2021qsw,Hristov:2022lcw,Hristov:2024cgj}, equivariant CY$_4$ volumes and topological strings \cite{Cassia:2025aus,Cassia:2025jkr}, and equivariant localization in quantum M-theory \cite{BenettiGenolini:2026cyc} --- yet in all of them the $N$-independent constant is left undetermined. The closed-forms of this paper supply precisely the missing data, and it is an interesting question whether they can be \emph{derived} within such a framework. 
\end{itemize}

\begin{acknowledgments}
We are grateful to Nikolay Bobev, Pieter-Jan De Smet, Morteza Hosseini, and Siyul Lee for valuable discussions. JH is supported by the National Research Foundation of Korea (NRF) grant funded by the Korean government (MSIT), Grant No.~RS-2024-00449284; by the Sogang University Research Grant No.~202410008.01; and by the Basic Science Research Program of the NRF funded by the Ministry of Education through the Center for Quantum Spacetime (CQUeST), Grant No.~RS-2020-NR049598.
\end{acknowledgments}

\appendix

\section{Large-$k$ expansions from the integral representations}\label{app:series}

In this appendix we derive the large-$k$ expansions quoted in \eqref{g0:closed}, \eqref{f0:closed} and \eqref{ADHMg0:closed} from the corresponding integral representations; read backward, the same chains are how those integral representations were reconstructed from the fitted general-order coefficients in Sections~\ref{sec:ABJM:new} and \ref{sec:ADHM:new}. Throughout we use the abbreviation $L(x)\equiv\log\big(1-e^{-kx}\big)$ together with the formulae \cite{Zwillinger:2007} 
\begin{align}
	\int_0^\infty\!dx\,x^ne^{-\mu x}&=\fft{n!}{\mu^{n+1}}\,,\quad \zeta(2n)=\fft{2^{2n-1}\pi^{2n}|B_{2n}|}{(2n)!}\,,\nn\\
	\tanh t&=\sum_{n=1}^\infty\fft{4^n(4^n-1)B_{2n}}{(2n)!}\,t^{2n-1}\,,\nn\\
	t\coth t&=\sum_{n=0}^\infty\fft{4^nB_{2n}}{(2n)!}\,t^{2n}\,,\nn\\
	\log\fft{\sinh t}{t}&=\sum_{n=1}^\infty\fft{4^nB_{2n}}{2n\,(2n)!}\,t^{2n}\,,\label{app:formulae}
\end{align}
which in particular give the Taylor expansions of the two kernels of interest,
\begin{align}
	\log\cosh x&=\sum_{n=1}^{\infty}\fft{4^n(4^n-1)B_{2n}}{2n\,(2n)!}\,x^{2n}\,,\label{app:kernels}\\
	G(x)&=\sum_{n=1}^{\infty}\Big(6\cdot4^n-2+\fft{2}{n}\Big)\fft{4^nB_{2n}}{(2n)!}\,x^{2n}\,,\nn
\end{align}
with $G(x)=12x\coth2x-2x\coth x-4+4\log\fft{\sinh x}{x}$.

\subsection{ABJM constants}\label{app:series:ABJM}

The large-$k$ expansion in \eqref{g0:closed} follows from the chain 
\begin{align}
	&\fft{k}{\pi^2}\int_0^\infty\!dx\,L(x)\log\cosh x \label{app:g0series}\\
	&=-\fft{k}{\pi^2}\sum_{m=1}^\infty\fft1m\int_0^\infty\!dx\,e^{-mkx}\log\cosh x\nn\\
	&=-\fft{1}{\pi^2}\sum_{n=1}^\infty\fft{4^n(4^n-1)B_{2n}}{2n}\,\zeta(2n+2)\,k^{-2n}\nn\\
	&=\sum_{n=1}^{\infty}\bigg(\fft{2\pi}{k}\bigg)^{2n}\fft{(-1)^n4^n(4^n-1)\,|B_{2n}B_{2n+2}|}{n\,(2n+2)!}\,,\nn
\end{align}
where we have used $B_{2n}=(-1)^{n+1}|B_{2n}|$ and the sum over $n$ is understood as an asymptotic large-$k$ expansion obtained by term-wise integration. The identical chain applied to the TTI kernel $G$ gives
\begin{align}
	&\fft{k}{\pi^2}\int_0^\infty\!dx\,L(x)\,G(x)\label{app:f0series}\\
	&=-\fft{2}{\pi^2}\sum_{n=1}^\infty\big(3\cdot16^n-4^n+\tfrac{4^n}{n}\big)B_{2n}\zeta(2n+2)k^{-2n}\nn\\
	&=\sum_{n=1}^{\infty}\bigg(\fft{2\pi}{k}\bigg)^{2n}\fft{(-1)^n4^{n+1}(3n\cdot4^n-n+1)|B_{2n}B_{2n+2}|}{n(2n+2)!}\,,\nn
\end{align}
which is the second line of \eqref{f0:closed}.

\subsection{ADHM constant}\label{app:series:ADHM}
By \eqref{app:kernels}, the two-level ADHM kernel of \eqref{ADHMg0:closed} carries the Taylor coefficient $\fft{(4^n-1)(2\cdot4^n+1)B_{2n}}{8n\,(2n)!}$ at order $x^{2n}$, so the chain \eqref{app:g0series} applied verbatim gives
\begin{align}
	&\fft{\Nf}{\pi^2}\int_0^\infty\!dx\,\log\big(1-e^{-\Nf x}\big)\label{app:ADHMg0series}\\
	&\kern5em\times\bigg[\fft12\log\cosh x+\fft14\log\cosh\fft x2\bigg]\nn\\
	&=\sum_{n=1}^{\infty}\bigg(\fft{2\pi}{\Nf}\bigg)^{2n}\fft{(-1)^n(4^n-1)(2\cdot4^n+1)\,|B_{2n}B_{2n+2}|}{4n\,(2n+2)!}\,,\nn
\end{align}
which is the second line of \eqref{ADHMg0:closed}.

\section{$A$-function representations}\label{app:Aform}

In this appendix we derive the $A$-function representations \eqref{g0:Aform} and \eqref{f0:Aform} of the ABJM constants from the integral representations \eqref{g0:closed} and \eqref{f0:closed}, and pin down the identification behind the closed-form \eqref{frakf0} of the constant $\mathfrak f_0$. We use the shorthand $L(x)=\log(1-e^{-kx})$ of Appendix~\ref{app:series}, the elementary integrals ($\Re[a]>0$)
\begin{align}
	\int_0^\infty dx\,\log\big(1-e^{-ax}\big)&=-\fft{\pi^2}{6a}\,,\nn\\
	\int_0^\infty dx\,x\log\big(1-e^{-ax}\big)&=-\fft{\zeta(3)}{a^2}\,,\label{app:elementary}
\end{align}
and the two integral families
\begin{align}
	D(a,b)&\equiv\int_0^\infty dx\,\log\big(1-e^{-ax}\big)\log\big(1-e^{-bx}\big)\,,\nn\\
	J(a,b)&\equiv\int_0^\infty dx\,\fft{x\log(1-e^{-ax})}{e^{bx}-1}\,,\label{app:DJ}
\end{align}
related by the exchange identity obtained from integration by parts,
\begin{align}
	b\,J(a,b)+a\,J(b,a)=-D(a,b)\,.\label{app:exchange}
\end{align}
The rescaling $x\to2x/a$ expresses $J$ through the integral representation \eqref{Afct} of the $A$-function,
\begin{align}
	J(a,b)&=\fft{4}{a^2}\,\tilde A\Big(\fft{2b}{a}\Big)\,,\nn\\
	\tilde A(q)&\equiv\fft{\pi^2}{q^2}\,A(q)-\fft{2\zeta(3)}{q^3}+\fft{\zeta(3)}{8}\,.\label{app:JtoA}
\end{align}
Note that $\tilde A(q)=J(2,q)$ is precisely the integral entering the first line of \eqref{Afct}.

\subsection{$A$-function form of $\hat g_0$}\label{app:Aform:g0}
Using the elementary decomposition of the kernel $\log\cosh x=x-\log2+\log(1-e^{-4x})-\log(1-e^{-2x})$, the identity \eqref{g0:Aform} follows from \eqref{g0:closed} in a single chain:
\begin{align}
	&\hat g_0(k,\Dsc)+\fft{\zeta(3)}{32\pi^2}k^2\label{app:g0Aform}\\
	&=-\fft16\log 2+\fft{k}{\pi^2}\int_0^\infty dx\,L(x)\log\cosh x
	\nn\\
	&=\fft{k}{\pi^2}\bigg[-\fft{\zeta(3)}{k^2}+D(k,4)-D(k,2)\bigg]\nn\\
	&=-\fft{\zeta(3)}{\pi^2k}-\fft{k}{\pi^2}\Big[4J(k,4)+kJ(4,k)-2J(k,2)-kJ(2,k)\Big]\nn\\
	&=-\fft{\zeta(3)}{\pi^2k}-\fft{1}{\pi^2}\bigg[\fft{16}{k}\,\tilde A\Big(\fft8k\Big)+\fft{k^2}{4}\,\tilde A\Big(\fft k2\Big)\nn\\
	&\kern8em-\fft{8}{k}\,\tilde A\Big(\fft4k\Big)-k^2\tilde A(k)\bigg]\nn\\
	&=A(k)-A\Big(\fft k2\Big)+\fft{k}{2}A\Big(\fft4k\Big)-\fft{k}{4}A\Big(\fft8k\Big)-\fft{3\zeta(3)}{32\pi^2}k^2\,,\nn
\end{align}
where the third equality eliminates $D$ in favor of $J$ via the exchange identity \eqref{app:exchange}, and the last two apply \eqref{app:JtoA}.

%

\subsection{$A$-function form of $\hat f_0$}\label{app:Aform:f0}
The $A$-function form \eqref{f0:Aform} and the constant \eqref{frakf0} follow simultaneously from the integral representation \eqref{f0:closed} in a single chain:
\begin{widetext}
\begin{align}
	\fft{k}{\pi^2}\int_0^\infty dx\,L(x)\,G(x)
	&=\fft{k}{\pi^2}\int_0^\infty dx\,L(x)\bigg[14x-4-4\log(2x)+\fft{24x}{e^{4x}-1}-\fft{4x}{e^{2x}-1}+4\log\big(1-e^{-2x}\big)\bigg]\nn\\
	&=-\fft{14\zeta(3)}{\pi^2k}+\fft23\log\fft{4\pi}{k}+8\zeta'(-1)+\fft{k}{\pi^2}\Big[24J(k,4)-4J(k,2)+4D(k,2)\Big]\nn\\
	&=-4\hat g_0(k,\Dsc)-\fft{\zeta(3)}{8\pi^2}k^2-\fft{18\zeta(3)}{\pi^2k}+\fft23\log\fft{4\pi}{k}+8\zeta'(-1)+\fft{k}{\pi^2}\Big[8J(k,4)-4J(k,2)-4kJ(4,k)\Big]\nn\\
	&=-4\hat g_0(k,\Dsc)-kA\Big(\fft4k\Big)+\fft k2A\Big(\fft8k\Big)-4A\Big(\fft k2\Big)+\fft{\zeta(3)}{8\pi^2}k^2+\fft23\log\fft{4\pi}{k}+8\zeta'(-1)\nn\\
	&=-6\hat g_0(k,\Dsc)+2A(k)-6A\Big(\fft k2\Big)-\fft{\zeta(3)}{8\pi^2}k^2+\fft23\log\fft{4\pi}{k}+8\zeta'(-1)\,.\label{app:f0closed}
\end{align}
\end{widetext}
The second equality uses \eqref{app:elementary} together with
\begin{align}
	&\fft{k}{\pi^2}\int_0^\infty dx\,L(x)\log(2x)\label{app:logint}\\
	&=\fft{k}{\pi^2}\,\partial_s\bigg[\int_0^\infty\!dx\,(2x)^{s-1}L(x)\bigg]_{s=1}\nn\\
	&=-\fft{k}{\pi^2}\,\partial_s\bigg[\fft{2^{s-1}\Gamma(s)\,\zeta(s+1)}{k^s}\bigg]_{s=1}\nn\\
	&=-\fft{1}{\pi^2}\Big[\big(\log2-\log k+\Gamma'(1)\big)\zeta(2)+\zeta'(2)\Big]\nn\\
	&=\fft16-2\zeta'(-1)-\fft16\log\fft{4\pi}{k}\,,\nn
\end{align}
where we have used the relation
\begin{align}
	\Gamma'(1)\,\zeta(2)+\zeta'(2)=\zeta(2)\big[\log2\pi-1+12\zeta'(-1)\big]\label{app:zetarel}
\end{align}
obtained as the $s$-derivative of the functional equation $\zeta(s)=2^s\pi^{s-1}\sin\fft{\pi s}{2}\,\Gamma(1-s)\,\zeta(1-s)$ at $s=-1$; the third employs the exchange identity \eqref{app:exchange} along with $\fft{k}{\pi^2}[D(k,4)-D(k,2)]=\hat g_0(k,\Dsc)+\fft{\zeta(3)}{32\pi^2}k^2+\fft{\zeta(3)}{\pi^2k}$; the fourth applies \eqref{app:JtoA}; and the fifth trades $kA(\fft4k)$ and $kA(\fft8k)$ for $\hat g_0$ and $A$-functions via \eqref{g0:Aform}. 

Note that adding the explicit terms of \eqref{f0:closed} to \eqref{app:f0closed} expresses $\hat f_0$ in terms of $\hat g_0$ and $A$-functions up to the single number $\mathfrak f_0+8\zeta'(-1)+\fft23\log4\pi$. Identifying its fitted value, within the numerical precision, as $-\fft52\log2$ --- motivated by its counterpart in $\hat g_0$ --- gives \eqref{frakf0}; the remaining terms then assemble into the 2nd expression of \eqref{f0:Aform}, and the exact values \eqref{f0:special} confirm the identification independently. Inserting \eqref{g0:Aform} finally gives the 1st expression in \eqref{f0:Aform}. 

\bibliography{M2-constants}

\end{document}